\documentclass[12pt]{article}

\usepackage{amsmath,amssymb,epsfig,subfigure,color,soul,fleqn,rotating}
\usepackage{ulem}
\usepackage{cite}
\usepackage[colorlinks,citecolor=blue,urlcolor=blue,linkcolor=blue]{hyperref}

\newcommand{\be}{\begin{eqnarray}}
\newcommand{\ee}{\end{eqnarray}}
\newcommand{\nee}{\nonumber\end{eqnarray}}
\newcommand{\nn}{\nonumber\\}

\newcommand{\drbar}{{\overline{\rm DR}}}

\newcommand{\msb}[1] {m_{\sb_{#1}}}
\newcommand{\mch}[1] {m_{\ti \x^+_{#1}}}
\newcommand{\mnt}[1] {m_{\ti \x^0_{#1}}}

\newcommand{\msg}    {m_{\ti g}}
\newcommand{\msu}[1] {m_{\ti u_{#1}}}

\def\msb          { \overline{\rm MS}}

\def\gev             {{\rm GeV}}

\newcommand{\gsim}{\;\raisebox{-0.9ex}
           {$\textstyle\stackrel{\textstyle >}{\sim}$}\;}

\def\be            {\begin{equation}}
\def\ee            {\end{equation}}
\def\bea            {\begin{eqnarray}}
\def\eea            {\end{eqnarray}}

\def\a              {\alpha}
\def\b               {\beta}
\def\d               {\delta}

\def\x               {\chi}
\def\ti              {\tilde}
\def\sq              {\ti q}
\def\st              {\ti t}

\def\ch              {\ti \x^\pm}

\def\nt              {\ti \x^0}
\def\sg              {\ti g}

\def\barc             {\bar{c}}
\def\su                {\ti{u}}
\def\sto                  {\ti{t}}
\def \sca                 {\ti{c}}

\def\dll            {\d^{LL}_{23}}
\def\durr            {\d^{uRR}_{23}}
\def\durl            {\d^{uRL}_{23}}
\def\dulr            {\d^{uLR}_{23}}

\newcommand{\thw}{\theta_{\textit{\tiny{W}}}}

\newcommand{\AddrVienna}{
\it Universit\"at Wien, Fakult\"at f\"ur Physik,
A-1090 Vienna, Austria \\}

\newcommand{\AddrGAKUGEI}{%
 \it Department of Physics, Tokyo Gakugei University, Koganei,
Tokyo 184-8501, Japan\\}

\newcommand{\AddrRIKEN}{%
 \it RIKEN Nishina Center for Accelerator-Based Science, 
 Wako, Saitama 351-0198, Japan\\}

\newcommand{\AddrHEPHY}{%
 \it Institut f\"ur Hochenergiephysik der \"Osterreichischen Akademie
der Wissenschaften, A-1050 Vienna, Austria\\}

\title{$h^0 \to c \bar{c}$ as a test case for quark flavor violation in the MSSM}

\author{A. Bartl${}^{1}$, H. Eberl${}^2$, E. Ginina${}^{2}$, K.~Hidaka${}^{3,4}$,\\
W.~Majerotto${}^2$}
\date{
\small  $^1$ \AddrVienna
        $^2$ \AddrHEPHY
        $^3$ \AddrGAKUGEI 
        $^4$ \AddrRIKEN}

\definecolor{darkgreen}{rgb}{0,.5,0}

\begin{document}

%**********
\begin{flushright}
HEPHY-PUB 942/14\\
UWThPh-2014-26\\
RIKEN-MP-101
\end{flushright}
\begingroup
\let\newpage\relax% Void the actions of \newpage
\maketitle
\endgroup
%**********

\maketitle
\thispagestyle{empty}

\begin{abstract}
We compute the decay width of $h^0 \to c \bar{c}$ in the MSSM with quark flavor 
violation (QFV) at full one-loop level adopting the $\overline{\rm DR}$ 
renormalization scheme.
We study the effects of $\ti{c}-\ti{t}$ mixing, taking into 
account the constraints from the B meson data. 
We show that the full one-loop corrected decay width $\Gamma (h^0 \to c \bar{c})$ 
is very sensitive to the MSSM QFV parameters.
In a scenario with large $\ti{c}_{L,R}-\ti{t}_{L,R}$ mixing $\Gamma (h^0 \to c \bar{c})$ can differ up to $\sim \pm 35\%$ from its SM value.
After estimating the uncertainties of the width, we conclude that an observation of these SUSY QFV effects is possible at an $e^+ e^-$ collider (ILC). 

%
%{\bf Keywords:} Phenomenology of the general MSSM, Non-minimal flavor violation, Collider Physics
%
%{\bf PACS:} 

%
\end{abstract}

\clearpage

\tableofcontents

\section{Introduction}

The properties of the Higgs boson, discovered at the LHC, CERN, with a mass 
of $125.15\pm0.24~\gev$ (averaged over the values given 
by ATLAS~\cite{Aad:2014aba, Kado} and CMS~\cite{CMSHiggs, David})~\cite{Ellis_LHCP2014}, are
consistent with the prediction of the Standard Model (SM)~\cite{LHCcrosssecs}. Future experiments at LHC at higher energy ($\sqrt{s}=14$ TeV) and higher luminosity will provide more precise data on Higgs boson observables, as Higgs production cross sections, decay branching ratios etc.. 
Even more precise data
can be expected at a future $e^+ e^-$ linear collider (ILC).
This will allow one to test the SM more accurately and will give information on physics beyond the SM. The discovered Higgs boson could also be the lightest neutral Higgs boson $h^0$ of the Minimal Supersymmetric Standard Model (MSSM)~\cite{LHCcrosssecs, Djouadi:2005gi}.

The decays of $h^0$ are usually assumed to be quark flavor conserving (QFC). However, quark flavor violation (QFV) in the squark sector may significantly influence the decay widths of $h^0$ at one-loop level. 
In particular, the rate 
of the $h^0$ decay into a charm-quark pair, $h^0 \to c \bar{c}$, may be 
significantly different from the SM prediction due to squark generation 
mixing, especially that between the second and the third squark generations 
($\tilde{c}_{L,R}-\st_{L,R}$ mixing). This possibility will be studied in 
detail in the present paper.

It is well known that the mixing between the first and the second squark generations is strongly suppressed by the data on K physics~\cite{pdg2014}. Therefore, we assume mixing between the second and the third squark generations, respecting the constraints from B physics.
In the MSSM this mixing was theoretically studied for squark and gluino production and decays at the LHC~\cite{Bozzi:2007me, Fuks:2008ab, Bartl:2010du, Bruhnke:2010rh, Hurth:2009ke, Bartl:2009au, Bartl:2011wq, Fuks:2011dg, Nomura:2007ap}. 

The outline of the paper is as follows: In Section~\ref{sec:sq.matrix} we shortly give the definitions of the QFV squark mixing parameters. In Section~\ref{sec:h2ccb} we present the calculation of the width of $h^0 \to c \bar{c}$ at full one-loop level in the $\drbar$ renormalization scheme with quark flavor violation within the MSSM. In particular, we give formulas for the important one-loop gluino contribution. In Section~\ref{sec:num} we present a detailed numerical analysis. 
In Section~\ref{sec:uncert} we study the feasibility of 
observing the SUSY QFV effects in the decay $h^0 \to c \bar{c}$ at ILC 
by estimating the theoretical uncertainties.
Section~\ref{sec:concl} contains our conclusions. 

%
%
%%%%%%%%%%%%%%%%%%%%%%%%%%%%%%%%%%%%%%%%%%%%%%%%%%%%%%%%%%%%%%%%%%%%%%%
\section{Definition of the QFV parameters}
\label{sec:sq.matrix}
%%%%%%%%%%%%%%%%%%%%%%%%%%%%%%%%%%%%%%%%%%%%%%%%%%%%%%%%%%%%%%%%%%%%%%

In the MSSM's super-CKM basis of $\sq_{0 \gamma} =
(\sq_{1 {\rm L}}, \sq_{2 {\rm L}}, \sq_{3 {\rm L}}$,
$\sq_{1 {\rm R}}, \sq_{2 {\rm R}}, \sq_{3 {\rm R}}),~\gamma = 1,...6,$  
with $(q_1, q_2, q_3)=(u, c, t),$ $(d, s, b)$, one can write the squark mass matrices in their most general $3\times3$-block form~\cite{Allanach:2008qq}
\begin{equation}
    {\cal M}^2_{\tilde{q}} = \left( \begin{array}{cc}
        {\cal M}^2_{\tilde{q},LL} & {\cal M}^2_{\tilde{q},LR} \\[2mm]
        {\cal M}^2_{\tilde{q},RL} & {\cal M}^2_{\tilde{q},RR} \end{array} \right),
 \label{EqMassMatrix}
\end{equation}
with $\tilde{q}=\tilde{u},\tilde{d}$. The left-left and right-right blocks in eq.~(\ref{EqMassMatrix}) are given by
\begin{eqnarray}
    & &{\cal M}^2_{\tilde{u},LL} = V_{\rm CKM} M_Q^2 V_{\rm CKM}^{\dag} + D_{\tilde{u},LL}{\bf 1} + \hat{m}^2_u, \nonumber \\
    & &{\cal M}^2_{\tilde{u},RR} = M_U^2 + D_{\tilde{u},RR}{\bf 1} + \hat{m}^2_u, \nonumber \\
    & & {\cal M}^2_{\tilde{d},LL} = M_Q^2 + D_{\tilde{d},LL}{\bf 1} + \hat{m}^2_d,  \nonumber \\
    & & {\cal M}^2_{\tilde{d},RR} = M_D^2 + D_{\tilde{d},RR}{\bf 1} + \hat{m}^2_d,
     \label{EqM2LLRR}
\end{eqnarray}
where $M_{Q,U,D}$ are the hermitian soft SUSY-breaking mass matrices of the squarks and
$\hat{m}_{u,d}$ are the diagonal mass matrices of the up-type and down-type quarks.
Furthermore, 
$D_{\tilde{q},LL} = \cos 2\beta m_Z^2 (T_3^q-e_q
\sin^2\theta_W)$ and $D_{\tilde{q},RR} = e_q \sin^2\theta_W \times$ $ \cos 2\beta m_Z^2$,
where
$T_3^q$ and $e_q$ are the isospin and
electric charge of the quarks (squarks), respectively, and $\theta_W$ is the weak mixing
angle.
Due to the $SU(2)_{\rm L}$ symmetry the left-left blocks of the up-type and down-type squarks in eq.~(\ref{EqM2LLRR}) are related
by the CKM matrix $V_{\rm CKM}$.
The left-right and right-left blocks of eq.~(\ref{EqMassMatrix}) are given by
\begin{eqnarray}
 {\cal M}^2_{\tilde{u},RL} = {\cal M}^{2\dag}_{\tilde{u},LR} &=&
\frac{v_2}{\sqrt{2}} T_U - \mu^* \hat{m}_u\cot\beta, \nonumber \\
 {\cal M}^2_{\tilde{d},RL} = {\cal M}^{2\dag}_{\tilde{d},LR} &=&
\frac{v_1}{\sqrt{2}} T_D - \mu^* \hat{m}_d\tan\beta,
\end{eqnarray}
where $T_{U,D}$ are the soft SUSY-breaking trilinear 
coupling matrices of the up-type and down-type squarks entering the Lagrangian 
${\cal L}_{int} \supset -(T_{U\alpha \beta} \ti{u}^\dagger _{R\a}\ti{u}_{L\b}H^0_2 $ 
$+ T_{D\alpha \beta} \ti{d}^\dagger _{R\a}\ti{d}_{L\b}H^0_1)$,
$\mu$ is the higgsino mass parameter, and $\tan\beta$ is the ratio of the vacuum expectation values of the neutral Higgs fields $v_2/v_1$, with $v_{1,2}=\sqrt{2} \left\langle H^0_{1,2} \right\rangle$.
The squark mass matrices are diagonalized by the $6\times6$ unitary matrices $U^{\tilde{q}}$,
$\tilde{q}=\tilde{u},\tilde{d}$, such that
\begin{eqnarray}
&&U^{\tilde{q}} {\cal M}^2_{\tilde{q}} (U^{\tilde{q} })^{\dag} = {\rm diag}(m_{\tilde{q}_1}^2,\dots,m_{\tilde{q}_6}^2)\,,
\end{eqnarray}
with $m_{\tilde{q}_1} < \dots < m_{\tilde{q}_6}$.
The physical mass eigenstates $\sq_i, i=1,...,6$ are given by $\sq_i =  U^{\sq}_{i \alpha} \sq_{0\alpha} $.

We define the QFV parameters in the up-type squark sector 
$\delta^{LL}_{\alpha\beta}$, $\delta^{uRR}_{\alpha\beta}$
and $\delta^{uRL}_{\alpha\beta}$ $(\alpha \neq \beta)$ as follows \cite{Gabbiani:1996hi}:
\begin{eqnarray}
\delta^{LL}_{\alpha\beta} & \equiv & M^2_{Q \alpha\beta} / \sqrt{M^2_{Q \alpha\alpha} M^2_{Q \beta\beta}}~,
\label{eq:InsLL}\\[3mm]
\delta^{uRR}_{\alpha\beta} &\equiv& M^2_{U \alpha\beta} / \sqrt{M^2_{U \alpha\alpha} M^2_{U \beta\beta}}~,
\label{eq:InsRR}\\[3mm]
\delta^{uRL}_{\alpha\beta} &\equiv& (v_2/\sqrt{2} ) T_{U\alpha \beta} / \sqrt{M^2_{U \alpha\alpha} M^2_{Q \beta\beta}}~,
\label{eq:InsRL}
\end{eqnarray}
where $\alpha,\beta=1,2,3 ~(\alpha \ne \beta)$ denote the quark flavors $u,c,t$.
In this study we consider $\ti{c}_R - \ti{t}_L$, 
$\ti{c}_L - \ti{t}_R$, $\ti{c}_R-\ti{t}_R$, and $\ti{c}_L - \ti{t}_L$ 
mixing which is described by the QFV parameters $\delta^{uRL}_{23}$, 
$\delta^{uLR}_{23} \equiv ( \delta^{uRL}_{32})^*$, $\delta^{uRR}_{23}$, and $\dll$, respectively.
We also consider $\ti{t}_L - \ti{t}_R$ mixing described by the QFC parameter $\delta^{uRL}_{33}$ which is defined 
by eq.~(\ref{eq:InsRL}) with $\alpha= \beta = 3$. 
All QFV parameters and $\delta^{uRL}_{33} $ are assumed to be real.

%********************************************************************
\section{$h^0 \to c \bar{c}$ at full one-loop level with flavor violation}
\label{sec:h2ccb}
%********************************************************************

\begin{figure*}[h!]
\centering
   { \mbox{ \resizebox{6.cm}{!}{\includegraphics{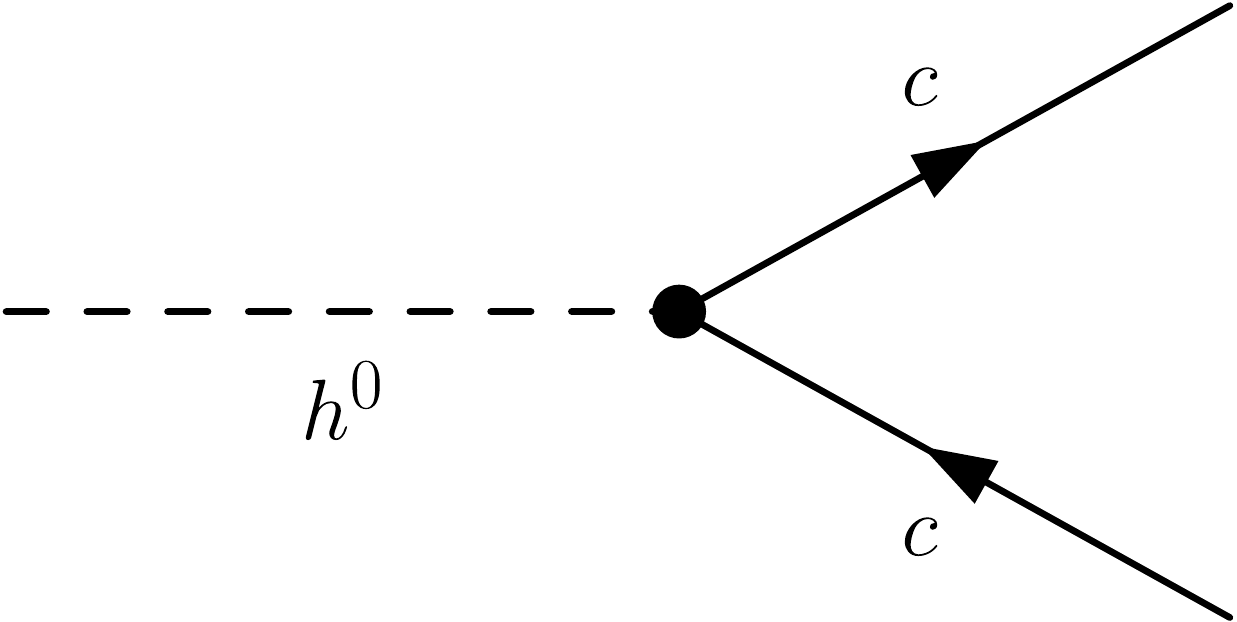}}}} %\hspace*{-0.8cm}}}
\caption{$h^0$ decay into a pair of charm quarks.}
   \label{h2ccbar}
\end{figure*}

We study the decay of the lightest neutral Higgs boson, $h^0$, into a pair of charm quarks (Figure~{\ref{h2ccbar}}) at full one-loop level in the general MSSM with quark flavor violation in the squark sector. The full one-loop decay width of $h^0 \to c \bar{c}$ was first calculated within the QFC MSSM by~\cite{Dabelstein:1995js}.
In \cite{Crivellin:2010er, Crivellin:2011jt, Crivellin:2012zz} higher order SUSY corrections for the Higgs-fermion-fermion vertices were calculated in the generic MSSM in an effective-field-theory approach.

The decay width of the reaction $h^0 \to c \bar{c}$ including one-loop contributions can be written as
\be
\Gamma(h^0 \to c \bar{c})=\Gamma^{\rm tree}(h^0 \to c \bar{c})+\delta \Gamma^{\rm  1loop}(h^0 \to c \bar{c})\,.
\label{decaywidth}
\ee
The tree-level decay width $\Gamma^{\rm tree}(h^0 \to c \bar{c})$ reads
\be
\Gamma^{\rm tree}(h^0 \to c \bar{c})=\frac{\rm N_C}{8 \pi} m_{h^0} (s_1^c)^2 \bigg( 1- \frac{4 m_c^2}{m_{h^0}^2}\bigg)^{3/2}\,, \quad {\rm with ~N_C}=3 \,,
\label{decaywidttree}
\ee
where $m_{h^0}$ is the on-shell (OS) mass of $h^0$ and the tree-level coupling $s_1^c$ is
\be
 s_1^c=-g \frac{m_c}{2 m_W} \frac{\cos{\a}}{\sin{\b}} =-\frac{h_c}{\sqrt{2}} \cos{\a}\,.
 \label{treecoup}
\ee
Here $\alpha$ is the mixing angle of the two CP-even Higgs bosons, $h^0$ and $H^0$~\cite{G&H}.

In the general MSSM at one-loop level, in addition to the diagrams that contribute within the SM, $\delta \Gamma^{\rm  1loop}(h^0 \to c \bar{c})$  also receives contributions from diagrams with additional Higgs bosons and supersymmetric particles. 
The contributions from SUSY particles are shown in Figure~\ref{1loopcontr}, neglecting the contributions from scalar leptons. The flavor violation is induced by one-loop diagrams with squarks that have a mixed quark flavor nature.  
In addition, the coupling of $h^0$ with two squarks $\su_i \su_j$ (see eq.~(\ref{coupluiujh}) of Appendix~\ref{sec:lag}) contains the trilinear coupling matrices $(T_U)_{ij}$ which for $i \ne j$ break quark flavor explicitly.

The one-loop contributions to $\Gamma(h^0 \to c \bar{c})$ contain three parts, QCD ($g$) corrections, SUSY-QCD ($\ti{g}$) corrections and electroweak (EW) corrections.  In the latter we also include the Higgs contributions. In the following we will mainly give details for the QCD and SUSY-QCD corrections. 

%***********************************************************************
\begin{figure*}
\centering
\subfigure[]{
   { \mbox{\hspace*{-1cm} \resizebox{6.cm}{!}{\includegraphics{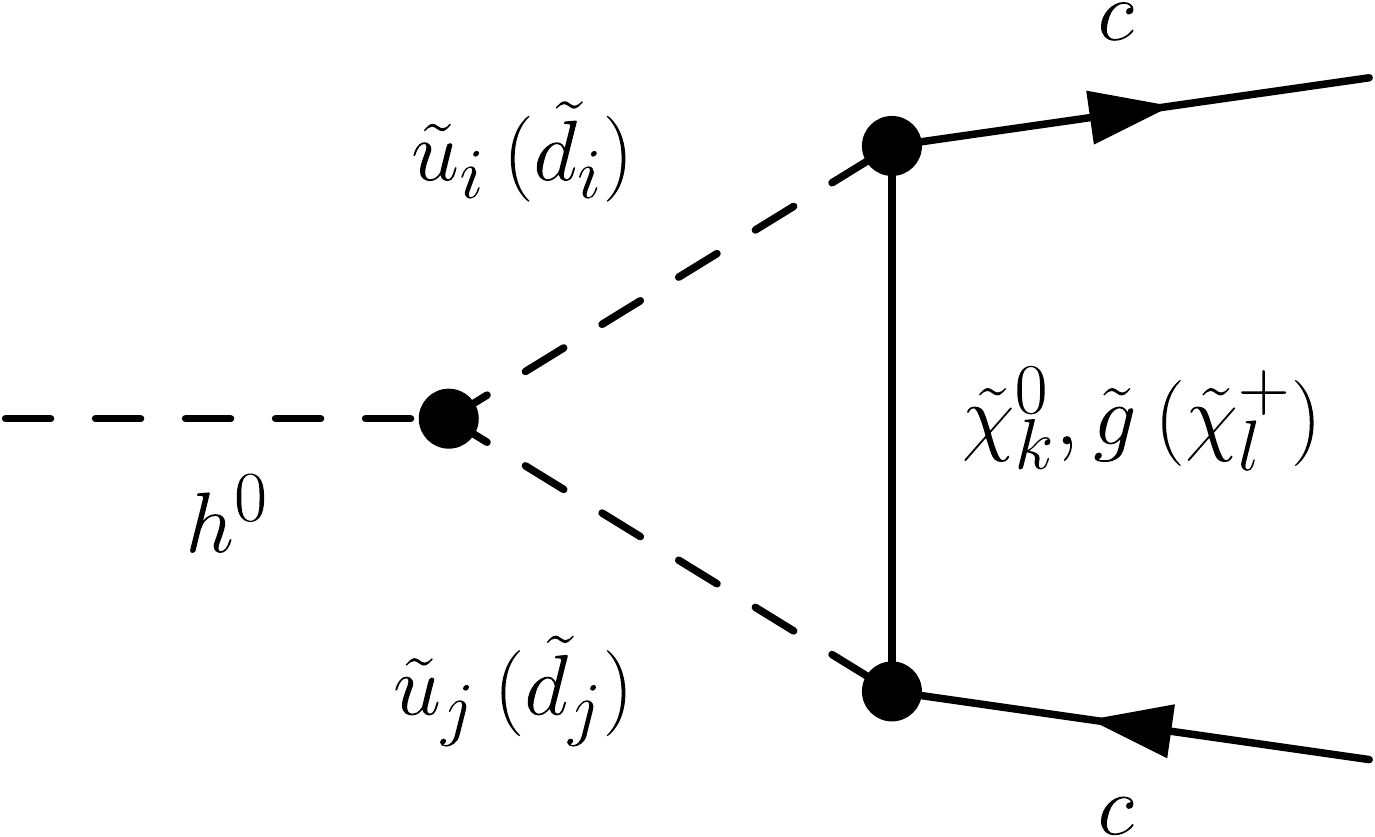}} \hspace*{-0.8cm}}}
   \label{diag1}} \qquad
 \subfigure[]{
   { \mbox{\hspace*{+0.cm} \resizebox{6.cm}{!}{\includegraphics{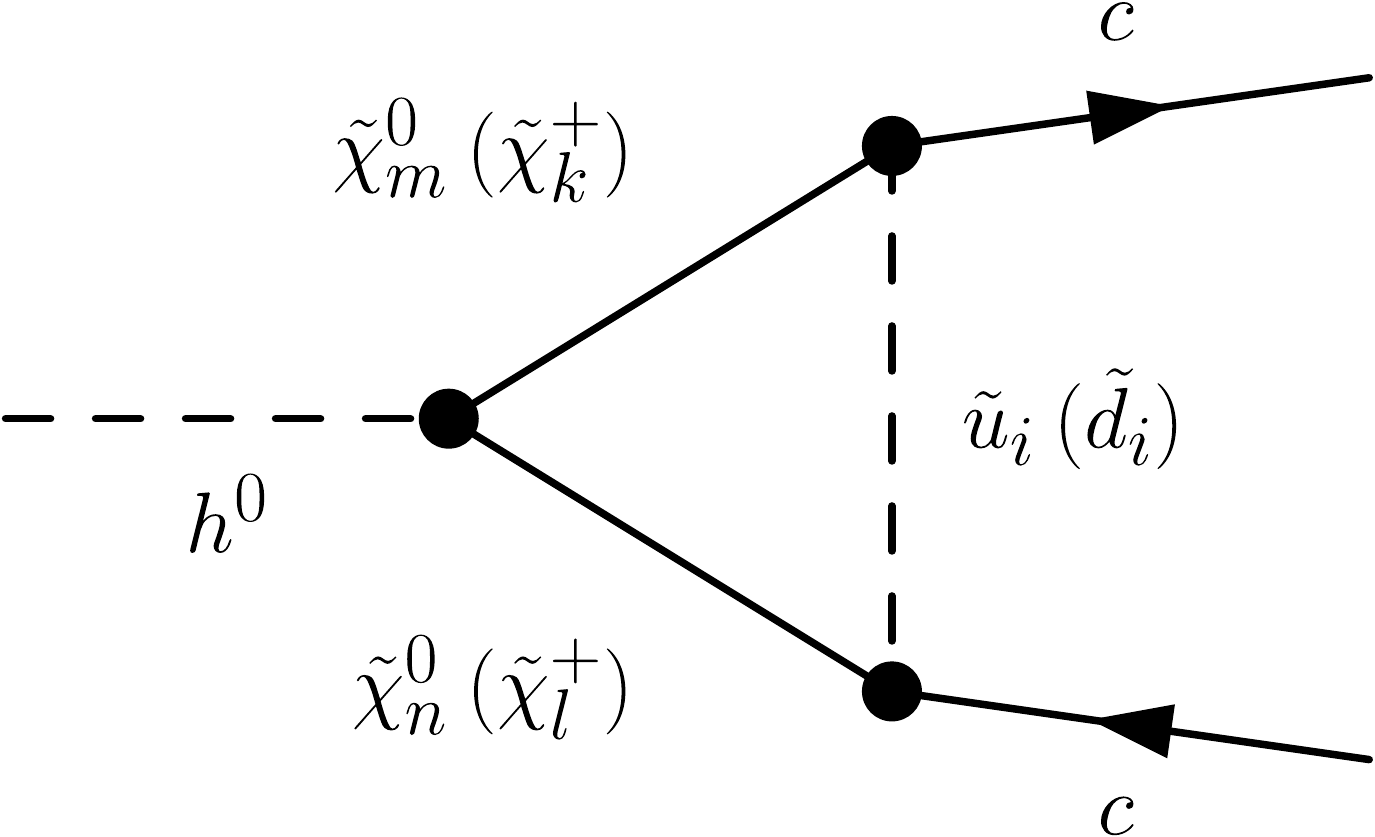}} \hspace*{-1cm}}}
  \label{diag2}
  }\\
 \subfigure[]{
   { \mbox{\hspace*{-1cm} \resizebox{6.cm}{!}{\includegraphics{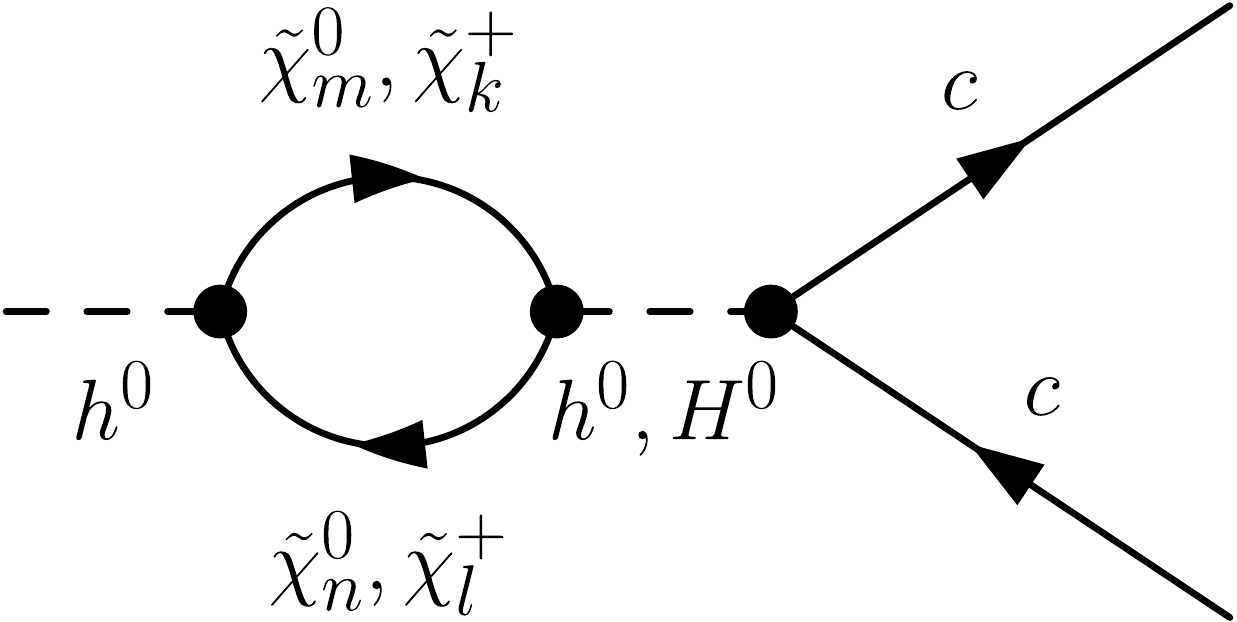}} \hspace*{-0.8cm}}}
   \label{diag3}} \qquad
 \subfigure[]{
   { \mbox{\hspace*{+0.cm} \resizebox{6.cm}{!}{\includegraphics{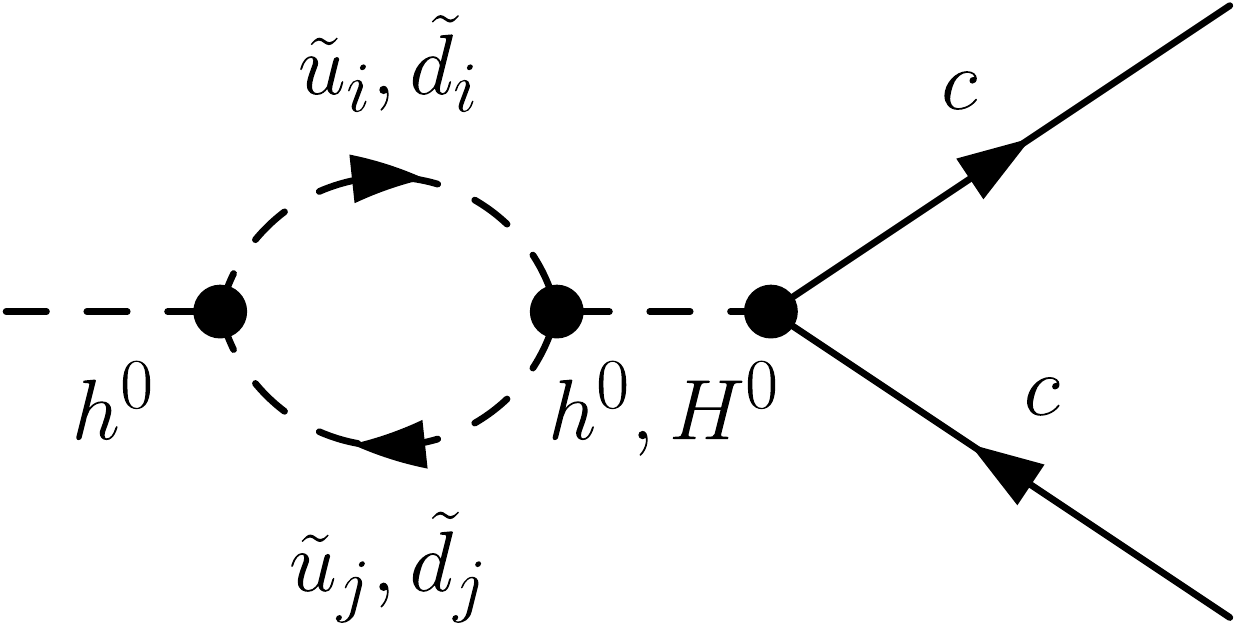}} \hspace*{-1cm}}}
  \label{diag4}
  } \\
  \subfigure[]{
   { \mbox{\hspace*{-1cm} \resizebox{6.cm}{!}{\includegraphics{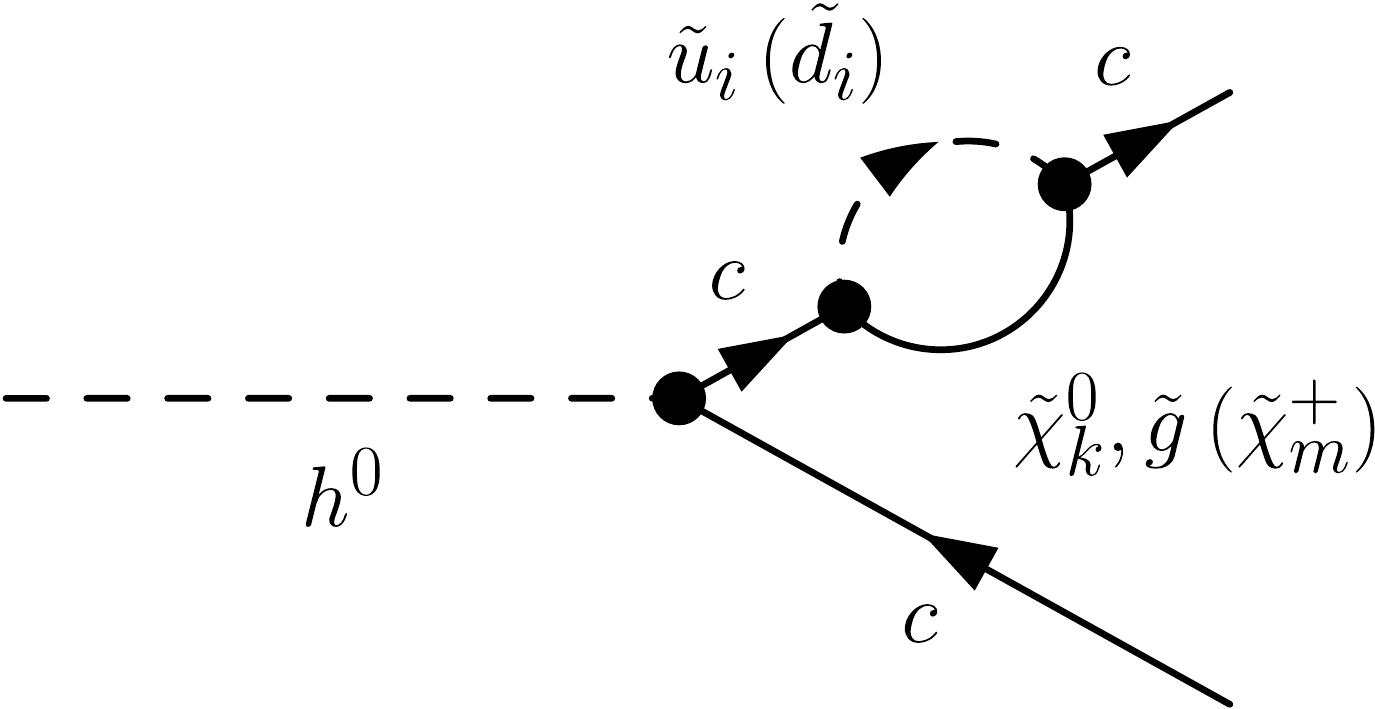}} \hspace*{-0.8cm}}}
   \label{diag5}} \qquad
 \subfigure[]{
   { \mbox{\hspace*{+0.cm} \resizebox{6.cm}{!}{\includegraphics{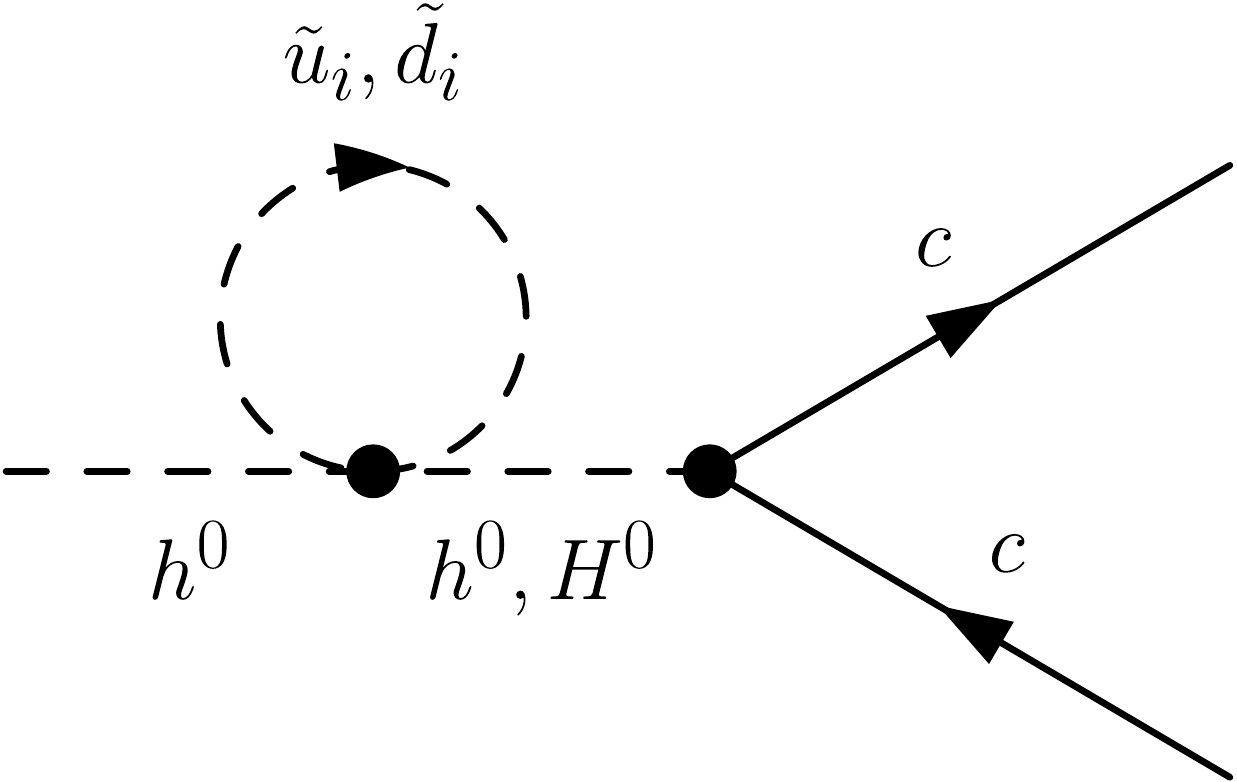}} \hspace*{-1cm}}}
  \label{diag6}
  } 
\caption{The main one-loop contributions with SUSY particles in $h^0 \to c \bar{c}$. The corresponding diagram to (e) with the self-energy contribution to the other charm quark is not shown explicitly. }
\label{1loopcontr}
\end{figure*}

%*************renormalization***************************************
\subsection{Renormalization procedure}
\label{sec:renorm}
%****************************************************

Loop calculations can lead to ultraviolet (UV) and infrared (IR) divergent result and therefore require renormalization. 
In order to get UV finite result we adopt in our study the $\drbar$ renormalization scheme, where all input parameters in the tree-level Lagrangian (masses, fields and coupling parameters) are UV finite, defined at the scale ${\rm Q}=125.5~\gev\simeq m_{h^0}$, and the UV divergence parameter $\Delta=\frac{2}{\epsilon}-\gamma+\ln 4 \pi$, where $\epsilon = 4-{\rm D}$ in a D-dimemsional space-time and $\gamma$ is the Euler-Mascheroni constant, is set to zero. 
The tree-level coupling is defined at the given scale and thus does 
not receive further finite shifts due to loop corrections. 
In order to obtain the shifts from the $\drbar$ masses and fields to the 
physical scale-independent masses and fields, we use on-shell renormalization conditions. 
To ensure IR convergence, we include in our calculations the contribution of the real hard gluon/photon radiation from the final charm quarks assuming a small gluon/photon mass~$\lambda$.

The one-loop corrected width of the process $h^0 \to c \bar{c}$ including hard gluon/photon radiation is given by 
\be
\Gamma (h^0 \to c \bar{c}) = \Gamma^{\rm tree}(h^0 \to c \bar{c})+\sum_{x=g, \ti{g}, {\rm EW}} \d \Gamma^x\,,
\label{totalres}
\ee
where $\d \Gamma^x$ read
\be
\d \Gamma^{\ti{g}} = \frac{3}{4 \pi} m_{h^0} s_1^c {\rm Re}(\d S_1^{c, \ti{g}}) \bigg( 1- \frac{4 m_c^2}{m_{h^0}^2}\bigg)^{3/2}\,,
\label{1loopcontr}
\ee
\be
\d \Gamma^{g/\rm EW} = \frac{3}{4 \pi} m_{h^0} s_1^c {\rm Re}(\d S_1^{c, g/\rm EW}) \bigg( 1- \frac{4 m_c^2}{m_{h^0}^2}\bigg)^{3/2}+ \Gamma^{\rm hard}(h^0 \to c \bar{c} g/\gamma)\,.
\label{1loopgEWcontr}
\ee
Note that all parameters in the tree-level coupling $s_1^c$, eq.~(\ref{treecoup}), are $\drbar$ running at the scale ${\rm Q}=125.5~\gev$.
The renormalized finite one-loop amplitude of the process is a sum of all vertex diagrams, the amplitudes arising from the wave-function renormalization constants and the amplitudes arising from the coupling counterterms.
Note that in the $\drbar$ renormalization scheme the counterterms contain only UV-divergent parts and have to cancel in order to yield a convergent result. The one-loop renormalized coupling correction can be written as
\be
\d S_1^{c, x} = \d S_1^{c, x (v)}+\d S_1^{c, x (w)}+\d S_1^{c, x (0)}, \quad x =g, \ti{g}, {\rm EW}\,,
\ee
where $\d S_1^{c, x (v)}$ is the vertex coupling correction, $\d S_1^{c, x (w)}$ is the wave-function coupling correction and $\d S_1^{c, x (0)}$ is the coupling counter term.
The tree-level interaction Lagrangian of the lightest Higgs boson $h^0$ and two charm quarks is given by eq.~(\ref{treelag}) in Appendix~\ref{sec:lag}. 
The renormalized Lagrangian ${\cal L}^{\rm ren}$ is obtained after making the replacement ${\cal L}^{\drbar}={\cal L}^{\rm ren} + \d {\cal L} $, where
 $\d  {\cal L} = - \d S_1^{c (v)} h^0 \barc c$ describes all vertex-type interactions.
 The coupling correction due to wave-function renormalization is given by
 \be
 \d S_1^{c (w)} = \frac{s_1^c}{2}\d Z_{h^0}+\frac{s_2^c}{2}\d Z_{h^0 H^0}+ \frac{s_1^c}{4}\left( \d Z^L_c + \d Z^{L*}_c+ \d Z^R_c + \d Z^{R*}_c\right)\,,
 \ee
where $s_2^c$ is the coupling of the heavier neutral Higgs $H^0$ and the charm quark, $s_2^c=-\frac{h_c}{\sqrt{2}} \sin{\a}$. 
The charm quark wave-function renormalization constants read
\begin{eqnarray}
\d Z_{c}^{L/R}&=&-{\rm \widetilde{Re}}~\Pi_{cc}^{L/R} (m_c)+\frac{1}{2 m_c}{\rm  \widetilde{Re}}\left(\Pi_{cc}^{S,~ L/R}(m_c)-
\Pi_{cc}^{S,~ R/L}(m_c)\right) \nonumber \\
&& -m_c{\rm  \widetilde{Re}}\bigg[ m_c\left(\dot{\Pi}_{cc}^{L/R}(m_c)+
\dot{\Pi}_{cc}^{R/L}(m_c)\right)\nonumber \\
&& +\dot{\Pi}_{cc}^{S,~ L/R}(m_c)+
\dot{\Pi}_{cc}^{S,~ R/L}(m_c) \bigg]\,,
\label{cquarckrenconst}
\end{eqnarray}
and the Higgs wave-function renormalization constants for the case of $h^0-H^0$ mixing are given by
\begin{equation}
\d Z_{h^0}=-{\rm  \widetilde{Re}}~\dot{\Pi}_{h^0 h^0}(m_{h^0}^{2})\,,
\label{hrenconst1}
\end{equation}
\begin{equation}
\d Z_{h^0 H^0}=\frac{2}{m_{h^0}^{2}-m_{H^0}^{2}}\left( {\rm  \widetilde{Re}}~\Pi_{h^0 H^0}(m_{h^0}^{2})-\d t_{h^0 H^0}\right)\,,
\label{hrenconst2}
\end{equation}
with the tadpole contribution
\begin{equation}
\d t_{h^0 H^0}= -\frac{1}{v}\bigg[ \tau_{h^0}\left(\frac{s_\a^2 c_\a}{c_\b}+ \frac{c_\a^2 s_\a}{s_\b}\right)+ 
\tau_{H^0}\left(-\frac{c_\a^2 s_\a}{c_\b}+ \frac{s_\a^2 c_\a}{s_\b}\right)\bigg]\,,
\end{equation}
where $c_\a=\cos \a$ and  $s_\a = \sin \a$. $\tau_{h^0}$ and $\tau_{H^0}$ are the loop corrections from the tadpole diagrams with $h^0$ and $H^0$, respectively.
In eqs.~(\ref{cquarckrenconst}), (\ref{hrenconst1}) and (\ref{hrenconst2}) $\widetilde{\rm Re}$ applied to the self-energies denoted by $\Pi$ takes the real part of the loop integrals,
 but leaves the possible complex couplings unaffected.
Finally, the coupling counter term $\d S_1^{c (0)}$ is given by
\be
\frac{\d S_1^{c (0)}} {s_1^c}= \left( \frac{\d g}{g}+\frac{\d m_c}{m_c}-\frac{\d m_W}{m_W}-\frac{\d \sin{\b}}{\sin{\b}}+\frac{\d \cos{\a}}{\cos{\a}}\right)_\Delta\,,
\label{delyuk}
\ee
where the subindex $\Delta$ means that only the part proportional to  the UV divergence parameter $\Delta$ is taken.
The explicit expressions for the shifts of the parameters in (\ref{delyuk}) can be found in \cite{wfrisch}. Note that $ \frac{\d g}{g} =\frac{\d e}{e}- \frac{\d \sin \theta_W}{\sin \theta_W}$ is used.

% 
%**********************************************************************
\subsection{One-loop gluon contribution}
\label{sec:gluon}
%**********************************************************************

The one-loop virtual gluon contribution to $\Gamma(h^0 \to c \bar{c})$ is given by
\be
\d \Gamma^{g}=\frac{3}{4 \pi} m_{h^0} s_1^c ~{\rm Re}(\d S_1^{c, g})\b^3\,,
\label{decaywidthgluon}
\ee
with $\b = (1-4m_c^2/m_{h^0}^2)^{1/2}$. $\d S_1^{c, g}$ contains terms originating from the vertex correction, the wave-function correction and the coupling correction due to gluon interaction,
\be
\d S_1^{c, g}=\d S_1^{c (g, v)}+\d S_1^{c (g, w)}+\d S_1^{c (g, 0)}\,.
\label{DS1cg}
\ee
The individual contributions in $\d S_1^{c, g}$ are given by
\be
\d S_1^{c (g, v)} = \frac{2\a_s}{3 \pi}  s_1^c \bigg[2 B_0-r-(m_{h^0}^2-2 m_c^2) C_0
-4 m_c^2C_1\bigg]\,,
\label{dsgv}
\ee
\be
\d S_1^{c (g, w)} = \frac{2 \a_s}{3 \pi}  s_1^c \bigg[- B_0- B_1+\frac{r}{2}+2 m_c^2(\dot{B_0}-\dot{B_1})\bigg] \,,
\ee
\be
\d S_1^{c (g, 0)} = \frac{2 \a_s}{3 \pi}  s_1^c \left( B_1-B_0+\frac{r}{2}\right)\,,
\label{dsg0}
\ee
where $r=0$ in the $\drbar$ scheme and $r=1$ in the ${\rm \overline{MS}}$ scheme. $B_k, \dot{B}_k$ and $C_k$ are the two- and three-point functions
\be
B_{k}=B_{k}(m_c^2, 0, m_c^2)\,,
\ee
\be
\dot{B}_{k}=\frac{\partial B_{k}(p^2,\lambda^2, m_c^2)}{\partial p^2}\bigg\vert_{p^2=m_c^2}\,, 
\ee
\be
C_{k}=C_{k}(m_c^2, m_{h^0}^2,  m_c^2,  \lambda^2, m_c^2,m_c^2)\,,
\ee
with $k=0,1$.
Summing up eqs.~(\ref{dsgv})-(\ref{dsg0}) one can write $\d S_1^{c, g}$ in the form
\be
\d S_1^{c, g} =  \frac{2}{3}\frac{\a_s}{ \pi} s_1^c \Delta^{{\rm H}, {\rm virt}} (\b)\,.
\label{delvirt}
\ee
Furthermore, we will use the result for the hard gluon radiation, given in Appendix~\ref{sec:brems}.
We can write eq.~(\ref{hgluon}) in the form
\be
\Gamma^{\rm hard}(h^0 \to c \bar{c} g)=\frac{3}{8 \pi} m_{h^0} (s_1^c)^2 \b^3  \frac{4}{3}\frac{ \a_s}{\pi} \Delta^{{\rm H}, {\rm hard}} (\b)\;.
\label{delhard}
\ee
Combining (\ref{decaywidthgluon}), (\ref{delvirt}) and (\ref{delhard}) for the gluon one-loop corrected convergent width we obtain
\be
\Gamma^{g}(h^0 \to c \bar{c})= \Gamma^{\rm tree}+\d \Gamma^{g}+\Gamma^{g, \rm hard}=\frac{3}{8 \pi} m_{h^0}(s_1^c)^2 \b^3 \left(1+ \frac{4}{3}\frac{ \a_s}{\pi} \Delta^{{\rm H}} (\b)\right)
\label{gamgluon}
\ee
where $\Delta^{\rm H} (\b) = \Delta^{{\rm H}, {\rm virt}} (\b)+\Delta^{{\rm H}, {\rm hard}} (\b)$ is the result of ~\cite{Braaten&Leveille} and its explicit expression can be found therein or e.g. in \cite{Dabelstein:1995js, M.Drees&Hikasa, Spira:1997dg}.
Eq.~({\ref{gamgluon}) can be written in a compact form as
\be
\Gamma^{g}(h^0 \to c \bar{c})=\Gamma^{\rm tree}(m_c\vert_{\rm OS})\left( 1+\frac{4}{3}\frac{\a_s}{\pi} \Delta^{\rm H} (\b) \right)\,,
\label{resg1}
\ee
where $m_c \vert_{\rm OS}$ denotes the on-shell (OS) charm quark mass. Note that the result for the photon one-loop corrected convergent width is obtained from 
(\ref{resg1}) by making the replacement $\frac{4}{3} \a_s \to e_c^2 \a$:
\be
\Gamma^{\gamma}(h^0 \to c \bar{c})=\Gamma^{\rm tree}(m_c\vert_{\rm OS})\left( 1+ \frac{4}{9}\frac{ \a }{ \pi} \Delta^H (\b) \right)\,,
\label{resg}
\ee
with $\a=e^2/(4 \pi)$.

For $m_c \ll m_{h^0}$ ($\b \to 1$)
\be
\Delta^{\rm H} = -3 \ln \frac{m_{h^0}}{m_c\vert_{\rm OS}} +\frac{9}{4}
\ee
and from eq.~(\ref{dsg0}) using eqs.~(\ref{b0m0m}) and (\ref{b1m0m}) we get
\be
\frac{\d m_c^g}{m_c}=\frac{\d S^{c,(g,0)}_1}{s_1^c}=\frac{\a_s}{3 \pi}\left( -6 \ln  \frac{m_{h^0}}{m_c\vert_{\rm OS}} + r -5 \right)\,.
\label{eq34}
\ee
For $\Gamma^{g}(h^0 \to c \bar{c})$ in the limit $m_c \ll m_{h^0}$ we obtain
\be
\Gamma^{g}(h^0 \to c \bar{c})=\Gamma^{\rm tree}(m_c|_{\rm SM})\left( 1+\frac{19 - 2r}{3}\frac{\a_s}{\pi} \right)\,,
\label{gamcor}
\ee
where in (\ref{gamcor}) we have absorbed
the logarithm of $\d m_c^g$ into
\be
m_c|_{\rm SM} = m_c\vert_{\rm OS} +\d m_c^g\,.
\label{eqmasses}
\ee
Combining eq.~(\ref{eq34}) with eqs.~(\ref{gamcor}) and (\ref{eqmasses}) one can see that the one-loop level $\Gamma^{g}(h^0 \to c \bar{c})$ does not depend on the parameter $r$.
In the numerical evaluation of $m_c\vert_{\rm SM}$ we follow the recipe given in~\cite{Eberl:1999he}, starting with eq.~(4) and we use $\a_s^{(2)}(\rm Q)$ given therein. 
In all other cases we take $\a_s(\rm Q)$ from SPheno~\cite{SPheno1, SPheno2}, where it is calculated at two-loop level within the MSSM.
In order to stay consistent, 
in our numerical calculations we have included in addition only the gluonic $\a_s^2$ contributions, taken from~\cite{Spira:1997dg}. 
With these, $\Gamma^{g}(h^0 \to c \bar{c})$ will be denoted as $\Gamma^{g, {\rm impr}}(h^0 \to c \bar{c})$,
\be
\Gamma^{g, {\rm impr}}(h^0 \to c \bar{c})=\Gamma^{\rm tree}(m_c|_{\rm SM})+\d \Gamma^g(m_c|_{\rm SM})\,.
\label{gamimpr}
\ee

%mymemo.tex

%**********************************************************************
\subsection{One-loop gluino contribution and decoupling limit}
%**********************************************************************

The one-loop gluino contribution to  $\Gamma(h^0 \to c \bar{c})$ , Fig.~\ref{sgvertex} and Fig.~\ref{sgself}, renormalised in the $\overline{\rm DR}$ scheme reads
\be
\d \Gamma^{\sg}=\frac{3}{4 \pi}m_{h^0}~ s_1^c ~{\rm Re}(\d S_1^{c, \sg})\b^3\,.
\label{decaywidthgluino}
\ee
$\d S_1^{c, \sg}$ acquires contributions from the vertex correction (Fig.~\ref{sgvertex}), the wave-function correction (Fig.~\ref{sgself}) and the coupling correction due to gluino interaction,
\be
\d S_1^{c, \sg}=\d S_1^{c (\sg, v)}+\d S_1^{c (\sg, w)}+\d S_1^{c (\sg, 0)}\,.
\label{DS1c}
\ee
In the following we will use the abbreviations
$\a_{ij} = U^{\su *}_{i2}U^{\su}_{j2}+U^{\su *}_{i5}U^{\su}_{j5}$  and $\b_{ij} = U^{\su *}_{i2}U^{\su}_{j5}+U^{\su *}_{i5}U^{\su}_{j2}$.
Note that applying Einstein sum convention we get $\a_{ii} = 2$ and $\b_{ii}=0$.  
Neglecting the charm quark mass and the Higgs boson mass compared to the squark and gluino masses, one can write the individual contributions as 
\bea
\d S_1^{c (\sg, v)} &=& \frac{\a_s}{3 \pi} \sum_{i,j=1}^6 G_{ij1}^{\su} \msg \b_{ij} C_0^{ij} \, ,
\label{vectorcorr}
\eea
\bea
\d S_1^{c (\sg, w)} &=& \frac{\a_s}{3 \pi} s_1^c \sum_{i=1}^6 \left( \a_{ii} B_1^{i} + 4 \msg \b_{ii} \dot{B}_0^{i}\right)\,,
\label{wfcsg}
\eea
where the coupling $G_{ij1}^{\su}$ is given in eq.~(\ref{coupluiujh}) of Appendix~\ref{sec:lag}.
For the following discussion of the gluino contribution in the large $m_{\ti{g}}$ limit we give the charm mass counter term $\d m_{c}^{\sg}$ in the OS scheme, which has a UV divergent and a finite contribution, 
\be
\d m_{c}^{\sg} = -\frac{\a_s}{3 \pi} \sum_{i=1}^6 \left( m_c \a_{ii} B_1^i + \msg \b_{ii}  B_0^i  \right)\,.
\label{dmcgl}
\ee
%%%%
\begin{figure*}[h!]
\centering
\subfigure[]{
   { \mbox{\hspace*{-1cm} \resizebox{6.3cm}{!}{\includegraphics{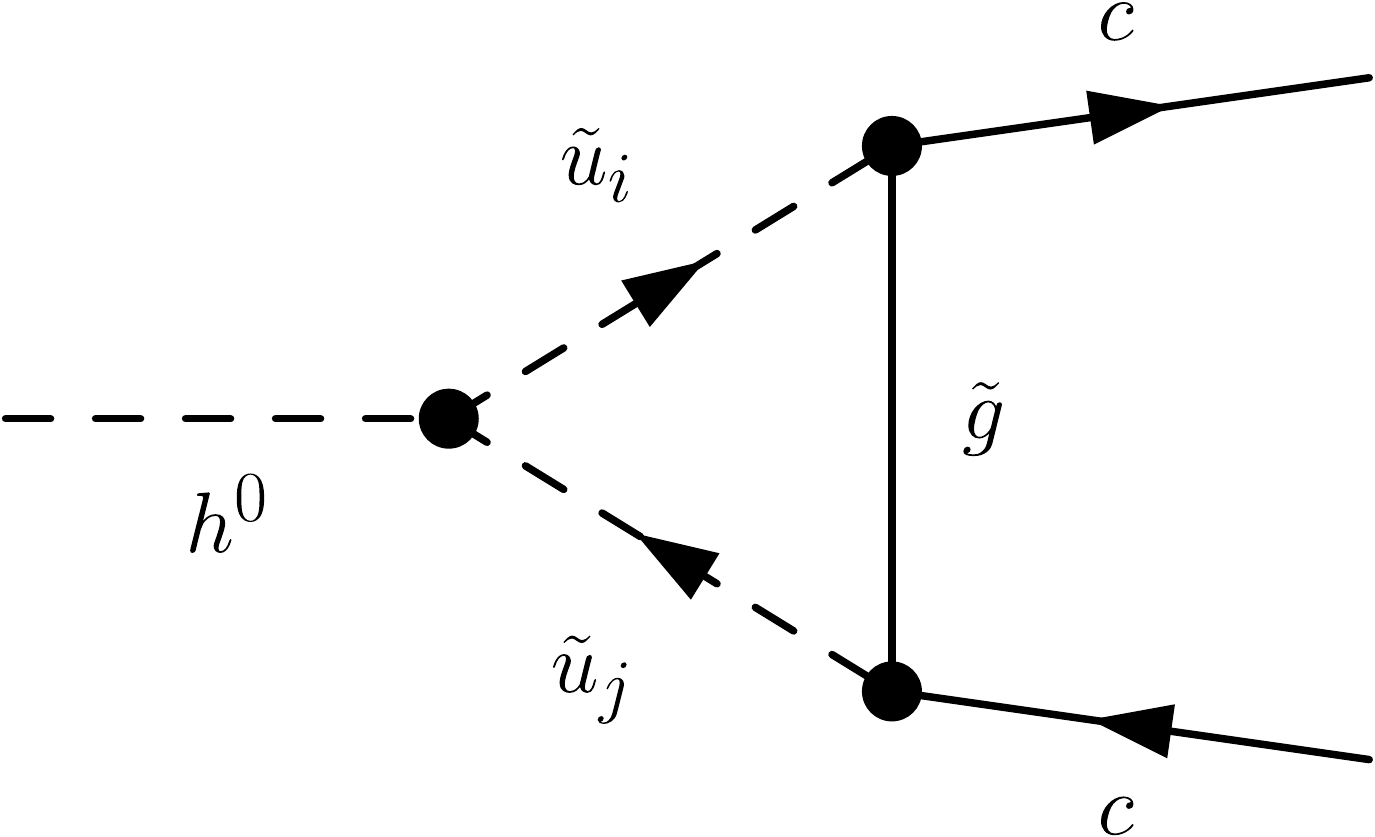}} %\hspace*{-0.8cm}
   }}
   \label{}}
 \subfigure[]{
   { \mbox{\hspace*{+0.5cm} \resizebox{7.3cm}{!}{\includegraphics{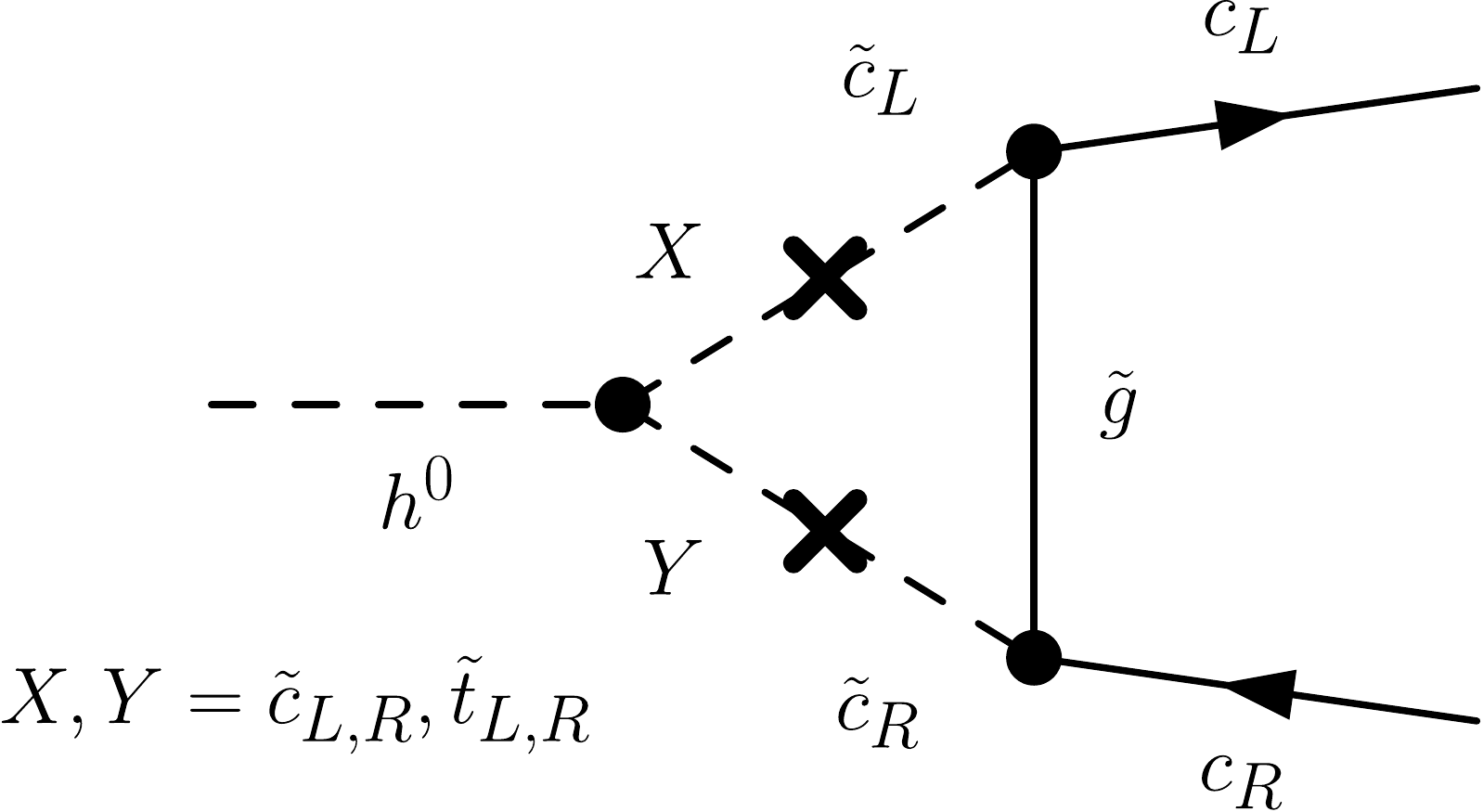}} \hspace*{-1cm}}}
  \label{}}
\caption{(a) Gluino vertex contribution to $h^0 \to c \bar{c}$ and (b) examples of quark flavor mixing in the gluino vertex contribution.}
\label{sgvertex}
\end{figure*}
\begin{figure*}[h!]
\centering
\subfigure[]{
   { \mbox{\hspace*{-1cm} \resizebox{6.3cm}{!}{\includegraphics{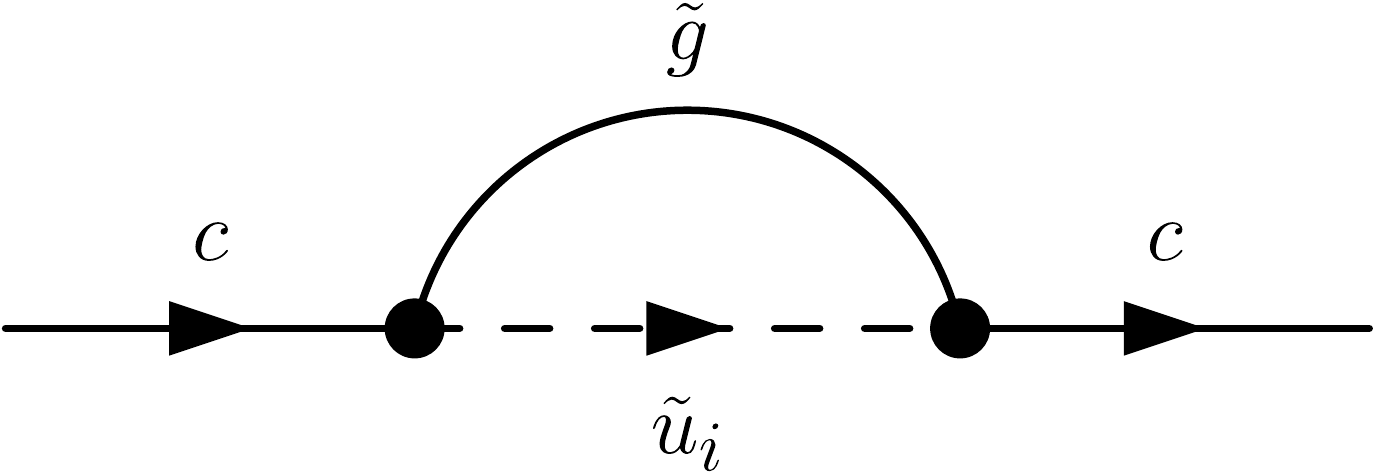}} %\hspace*{-0.8cm}
   }}
   \label{sgselfa}}
 \subfigure[]{
   { \mbox{\hspace*{+0.5cm} \resizebox{6.3cm}{!}{\includegraphics{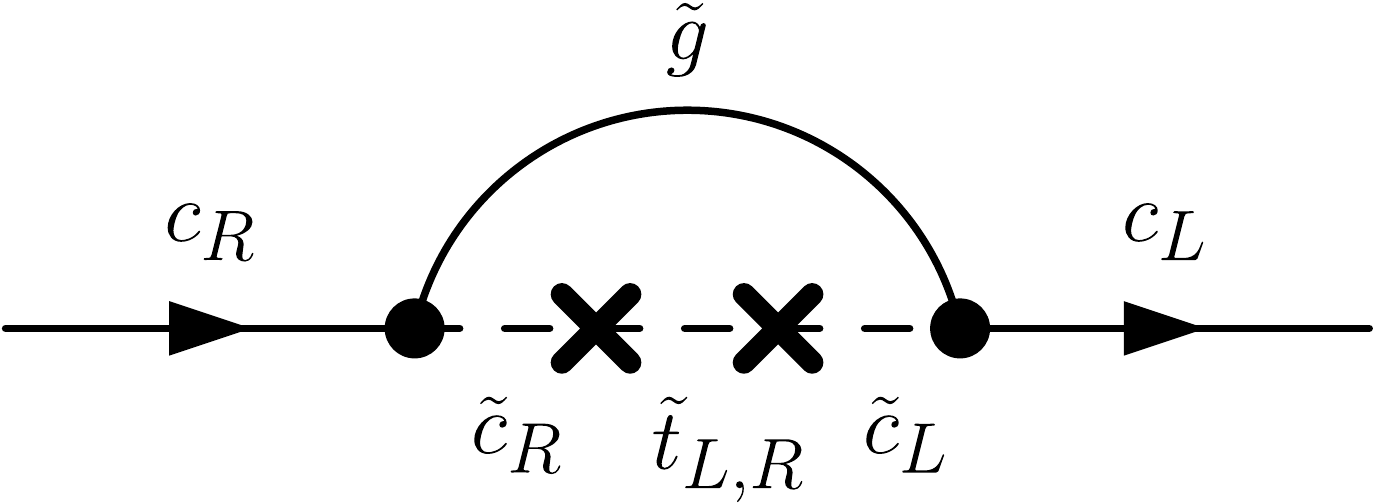}} \hspace*{-1cm}}}
  \label{sgselfb}}
\caption{(a) The gluino contribution to the charm quark self-energy and (b) examples of quark flavor mixing in the charm quark self-energy contribution with gluino.}
\label{sgself}
\end{figure*}
%%%%
For the gluino contribution we have $\d S_1^{c (\sg, 0)}/s_1^c   =  \d m_{c}^{\sg}/m_c$. Therefore, with eq.~(\ref{wfcsg}) we get 
\bea
\d S_1^{c (\sg, 0)} &=& -\frac{\a_s}{3 \pi}  s_1^c \sum_{i=1}^6 \left(\a_{ii} B_1^i + \frac{\msg}{m_c} \b_{ii}  B_0^i  \right)\,.
\label{dS1c-OS}
\eea
In the $\drbar$~scheme we need only the UV divergent part of (\ref{dS1c-OS}) which is
\bea
\d S_1^{c (\sg, 0)} &=& 6 \frac{\a_s}{3 \pi} s_1^c \Delta\, .
\eea
$\Delta$ is the UV divergence factor. In eqs.~(\ref{vectorcorr})-(\ref{dS1c-OS}) 
$B_{k}^i, \dot{B}_{0}^i$ and $C_{0}^i$ are the two- and three-point functions
\be
B_{k}^i=B_{k}(0,\msg^2, \msu i^2), \quad k=0,1,~ i=1,...,6,
\ee
\be
\dot{B}_{0}^i=\frac{\partial B_{0}(p^2, \msg^2, \msu i^2)}{\partial p^2}\bigg\vert_{p^2=0}\,,~ i=1,...,6,
\ee
\be
C_{0}^{ij}=C_{k}(0, 0, 0, \msg^2, \msu i^2,\msu j^2), \quad  i=1,...,6.
\ee
The total correction $\d S^{c, \sg}_1$ (eq.~(\ref{DS1c})) is given by
\bea
\hspace*{-1cm}
\overline{\rm DR}\,{\rm scheme:}\quad &
\d S_1^{c, \sg} & = \frac{\a_s}{3 \pi} \sum_{i,j=1}^{6}\bigg\{ \msg \b_{ij} \left( G_{ij1}^{\su} C_0^{ij}  + 4 s_1^c \d_{ij}  \dot{B}_0^{i}\right)
+ s_1^c  \d_{ij} \left( \a_{ii} B_1^{i} + \Delta\right) \bigg\}
\label{DS1c-DRbar}\\
\hspace*{-1cm}
{\rm OS~scheme:}\quad &
\d S_1^{c, \sg} & = \frac{\a_s}{3 \pi} \sum_{i,j=1}^{6}\bigg\{ \msg \b_{ij} \left( G_{ij1}^{\su} C_0^{ij}  + 4 s_1^c   \d_{ij}  \dot{B}_0^{i}\right)
- s_1^c  \d_{ij} \frac{\msg}{m_c} \b_{ii}  B_0^i  \bigg\}\,.
\label{DS1c-OS}\
\eea
As $B_1^{i} \to -\Delta/2$ and thus $ \a_{ii} B_1^{i}  \to -\Delta$, (\ref{DS1c-DRbar}) is UV convergent.
As $ \b_{ii} B_0^{i} \to 0$, also  (\ref{DS1c-OS}) is UV convergent. 

In the limit $\msg \to \infty$, from (\ref{C0simplxy}) it follows $\msg C_0^{ij} \to 0$ and from (\ref{dB0simpl}) it follows $\dot{B}_0^{i} \to 0$.
However, in this limit (\ref{B0simpl}) and (\ref{B1simpl}) 
become independent of the index~$i$ and
grow with $\ln \frac{\msg^2}{m_{h^0}^2}$. Therefore, $\b_{ii}  B_0^i \to 0$ guarantees decoupling of the gluino loop contribution 
in the OS~scheme.

In the $\drbar$ scheme for $\msg \to \infty$, we get
\be
\d S_1^{c, \sg} \sim \frac{2 \a_s}{3 \pi} s_1^c B_1^i \quad $\rm with$ \quad B_1\sim \ln \frac{\msg^2}{m_{h^0}^2}\,.
\ee 
At first sight it seems that the gluino contribution does not decouple for $\msg \to \infty$. However, the tree-level coupling $s_1^c$ (eq.~(\ref{treecoup})) contains a factor $m_c$. We have
\be
m_c (m_{h^0}) \vert_{\drbar} = m_c (m_c) \vert_{\overline{\rm MS}} +\d m_c^{\tilde{g}}+\ldots\,,
\ee
where we take $ m_c (m_c) \vert_{\overline{\rm MS}} =1.275~\gev$ as input~\cite{AguilarSaavedra:2005pw}.
$\d m_c^{\tilde{g}}$ is due to the self-energy contributions with gluino (see Figs.~\ref{sgselfa} and \ref{sgselfb}). 
We get
\be
\d m_{c}^{\sg}  \sim -\frac{2 \a_s}{3 \pi} m_c B_1^i\,.
\ee
Thus the sum $\Gamma^{\rm tree}+ \d \Gamma^{\tilde{g}}$ is indeed decoupling for $\msg \to \infty$. Analogously, this also holds for the chargino and neutralino contributions.

%**********************************************************************
\subsection{Total result for the width at full one-loop level}
%**********************************************************************

Finally, we want to sum up all contributions to get the total result for $\Gamma (h^0 \to c \bar{c})$ at full one loop level.

The one-loop result including gluino and EW contributions reads
\be
\Gamma^{\sg+{\rm EW}} (h^0 \to c \bar{c}) = \Gamma^{\rm tree} (m_c)+ \d \Gamma^{\tilde{g}}(m_c) + \d \Gamma^{{\rm EW}}(m_c)\,,
\label{gandEW}
\ee
where $\Gamma^{\rm tree}$, $\d \Gamma^{\tilde{g}} $ and $\d \Gamma^{{\rm EW}}$ are given by eqs.~(\ref{decaywidttree}), (\ref{decaywidthgluino}) and (\ref{1loopgEWcontr}), respectively. Note that eq.~(\ref{gandEW}) is a series expansion around $\Gamma^{\rm tree}(m_c)=\Gamma^{\rm tree}\left(m_c(m_{h^0})\vert_\drbar\right)$.\\
However, the improved result with gluon contribution (eq.~(\ref{gamimpr})) given by
\be
\Gamma (h^0 \to c \bar{c})^{g, {\rm impr}} = \Gamma^{\rm tree}(m_c\vert_{\rm SM})+ \d \Gamma^{g} (m_c\vert_{\rm SM})
\label{gluino}
\ee
is a series expansion around $ \Gamma^{\rm tree}(m_c\vert_{\rm SM})$.
In order to combine eqs.~(\ref{gandEW}) and (\ref{gluino}) in a consistent way we 
write: 
 \be
 \Gamma^{\rm tree}(m_c\vert_{\rm SM})= \Gamma^{\rm tree}(m_c)\frac{m_c^2\vert_{\rm SM}}{m_c^2}\,,
 \ee
 and therefore
\be
\Gamma^{\rm tree}(m_c\vert_{\rm SM})= \Gamma^{\rm tree}(m_c)- \Gamma^{\rm tree} (m_c) \frac{m_c^2-m_c^2\vert_{\rm SM}}{m_c^2}\,.
\ee
Thus, our total result can be written in the form
\be
\Gamma (h^0 \to c \bar{c}) \equiv \Gamma^{\rm impr} (h^0 \to c \bar{c}) = \Gamma^{\rm tree}(m_c)+\d \widetilde{\Gamma}^g + \d \Gamma^{\tilde{g}} + \d \Gamma^{\rm EW}\,,
\label{fullres}
\ee
where the new gluon contribution $\d \ti{\Gamma}^g $ is given by
\be
\d \widetilde{\Gamma}^g =  \d \Gamma^{g} (m_c\vert_{\rm SM})-\Gamma^{\rm tree}(m_c) \frac{m_c^2-m_c^2\vert_{\rm SM}}{m_c^2}\,.
\ee

%**********************************************************************
\section{Numerical results }
\label{sec:num}
%**********************************************************************

In order to demonstrate clearly the effect of QFV in the MSSM, we have explicitly chosen a reference scenario with a rather strong $\ti{c}-\ti{t}$ mixing. The MSSM parameters at $Q = 125.5~\gev \simeq m_{h^0}$ are given in Table~\ref{basicparam}. 
%
%%*****************************************************
\begin{table}[h!]
\caption{Reference QFV scenario: shown are the basic MSSM parameters 
at $Q = 125.5~{\rm GeV} \simeq m_{h^0}$,
except for $m_{A^0}$ which is the pole mass (i.e.\ the physical mass) of $A^0$, 
with $T_{U33} = - 2050$~GeV (corresponding to $\delta^{uRL}_{33} = - 0.2$). All other squark 
parameters not shown here are zero. }
\begin{center}
\begin{tabular}{|c|c|c|}
  \hline
 $M_1$ & $M_2$ & $M_3$ \\
 \hline \hline
 250~\gev  &  500~\gev &  1500~\gev \\
  \hline
\end{tabular}
\vskip0.4cm
\begin{tabular}{|c|c|c|}
  \hline
 $\mu$ & $\tan \beta$ & $m_{A^0}$ \\
 \hline \hline
 2000~\gev & 20 &  1500~\gev \\
  \hline
\end{tabular}
\vskip0.4cm
\begin{tabular}{|c|c|c|c|}
  \hline
   & $\alpha = 1$ & $\alpha= 2$ & $\alpha = 3$ \\
  \hline \hline
   $M_{Q \alpha \alpha}^2$ & $(2400)^2~\gev^2$ &  $(2360)^2~\gev^2$  & $(1850)^2~\gev^2$ \\
   \hline
   $M_{U \alpha \alpha}^2$ & $(2380)^2~\gev^2$ & $(1050)^2~\gev^2$ & $(950)^2~\gev^2$ \\
   \hline
   $M_{D \alpha \alpha}^2$ & $(2380)^2~\gev^2$ & $(2340)^2~\gev^2$ &  $(2300)^2~\gev^2$  \\
   \hline
\end{tabular}
\vskip0.4cm
\begin{tabular}{|c|c|c|c|}
  \hline
   $\delta^{LL}_{23}$ & $\delta^{uRR}_{23}$  &  $\delta^{uRL}_{23}$ & $\delta^{uLR}_{23}$\\
  \hline \hline
   0.05 & 0.2 &  0.03   &  0.06  \\
    \hline
\end{tabular}
\end{center}
\label{basicparam}
\end{table}
%
%****************************************************
\begin{table}[h!]
\caption{Physical masses in GeV of the particles for the scenario of Table~\ref{basicparam}.}
\begin{center}
\begin{tabular}{|c|c|c|c|c|c|}
  \hline
  $\mnt{1}$ & $\mnt{2}$ & $\mnt{3}$ & $\mnt{4}$ & $\mch{1}$ & $\mch{2}$ \\
  \hline \hline
  $260$ & $534$ & $2020$ & $2021$ & $534$ & $2022$ \\
  \hline
\end{tabular}
\vskip 0.4cm
\begin{tabular}{|c|c|c|c|c|}
  \hline
  $m_{h^0}$ & $m_{H^0}$ & $m_{A^0}$ & $m_{H^+}$ \\
  \hline \hline
  $126.08$  & $1498$ & $1500$ & $ 1501$ \\
  \hline
\end{tabular}
\vskip 0.4cm
\begin{tabular}{|c|c|c|c|c|c|c|}
  \hline
  $\msg$ & $\msu{1}$ & $\msu{2}$ & $\msu{3}$ & $\msu{4}$ & $\msu{5}$ & $\msu{6}$ \\
  \hline \hline
  $1473$ & $756$ & $965$ & $1800$ & $2298$ & $2301$ & $2332$ \\
  \hline
\end{tabular}
%*********************
\end{center}
\label{physmasses}
\end{table}
%
%****************************************************
\begin{table}[h!]
\caption{Flavor decomposition of $\su_1$ and $\su_2$ for the scenario of Table~\ref{basicparam}. Shown are the squared coefficients. }
\begin{center}
\begin{tabular}{|c|c|c|c|c|c|c|c|}
  \hline
  & $\su_L$ & $\sca_L$ & $\sto_L$ & $\su_R$ & $\sca_R$ & $\sto_R$ \\
  \hline \hline
 $\su_1$  & $0$ & $0.0004$ & $0.012$ & $0$ & $0.519$ & $0.468$ \\
  \hline 
  $\su_2$  & $0$ & $0.0004$ & $0.009$ & $0$ & $0.480$ & $0.509$ \\
  \hline
\end{tabular}
%*********************
\end{center}
\label{flavourdecomp}
\end{table}
%}
%****************************************************
%&&&&&&&&&&&&&&&&&&&&&&&&&&&&&&&&&&&&&PLOTS&&&&&&&&&&&&&&&&&&&&&&&&&&&&&&&&&&&&&&
%***********************************************************************
\begin{figure*}[h!]
\centering
\subfigure[]{
   { \mbox{\hspace*{-1cm} \resizebox{8.cm}{!}{\includegraphics{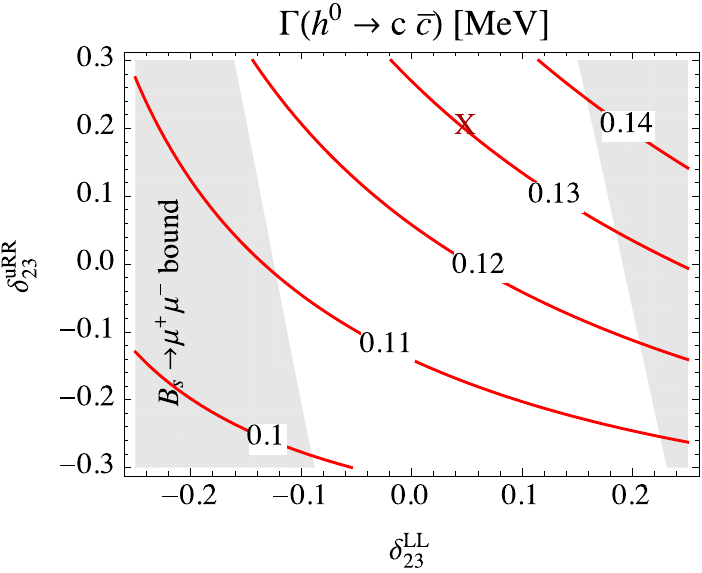}} \hspace*{-0.8cm}}}
   \label{fig1a}}
 \subfigure[]{
   { \mbox{\hspace*{+0.cm} \resizebox{8.cm}{!}{\includegraphics{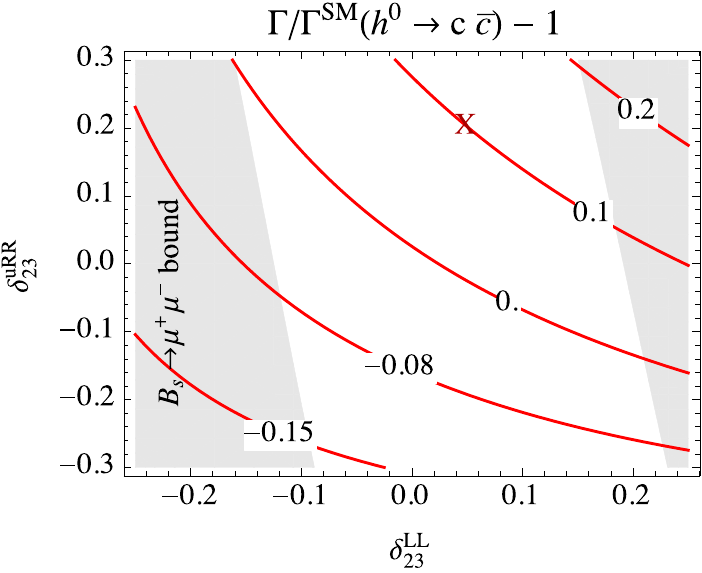}} \hspace*{-1cm}}}
  \label{fig1b}}\\
  \centering
  \subfigure[]{
   { \mbox{\hspace*{-1cm} \resizebox{8.cm}{!}{\includegraphics{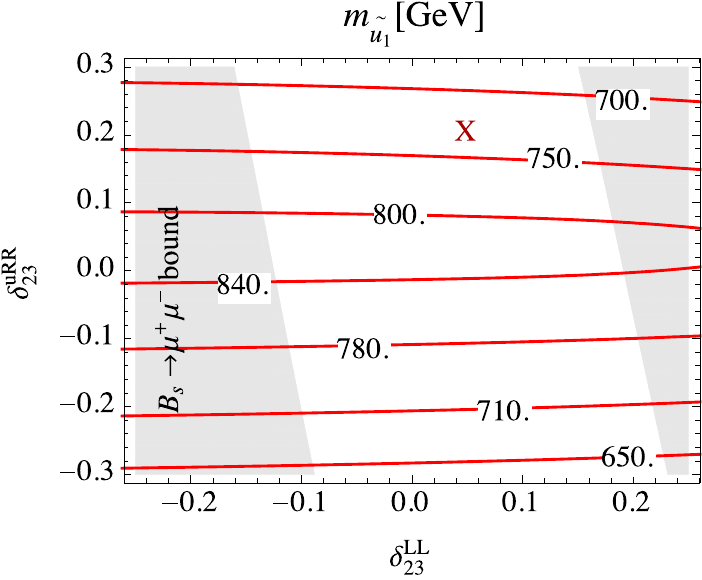}} \hspace*{-0.8cm}}}
   \label{fig1c}}
\caption{Dependence on the QFV parameters $\dll$ and $\durr$ of the width (a)~$\Gamma(h^0 \to c \bar{c})$
in MeV, (b) $\Gamma(h^0 \to c \bar{c})$/$\Gamma^{\rm SM}(h^0 \to c \bar{c})$ and (c) the mass of the lightest 
squark $\su_1$ in GeV. The gray region is excluded by the constraint from the B($B_s \to \mu^+ \mu^-$) data.
}
\label{fig1}
\end{figure*}
%
%***********************************************************************
\begin{figure*}[h!]
\centering
\subfigure[]{
   { \mbox{\hspace*{-1cm} \resizebox{8.cm}{!}{\includegraphics{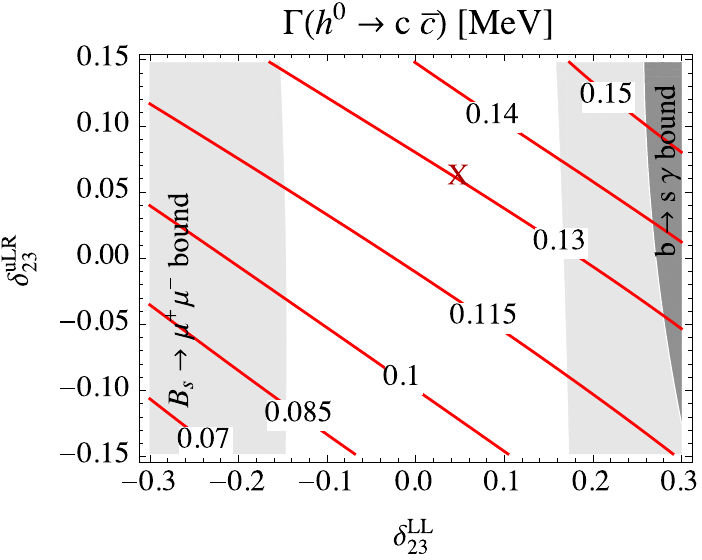}} \hspace*{-0.8cm}}}
   \label{fig2a}}
 \subfigure[]{
   { \mbox{\hspace*{+0.cm} \resizebox{8.cm}{!}{\includegraphics{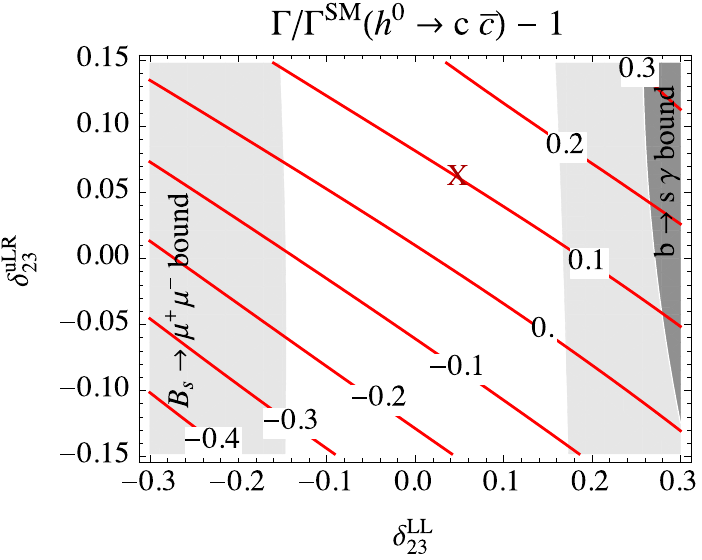}} \hspace*{-1cm}}}
  \label{fig2b}}\\
  \centering
  \subfigure[]{
   { \mbox{\hspace*{-1cm} \resizebox{8.cm}{!}{\includegraphics{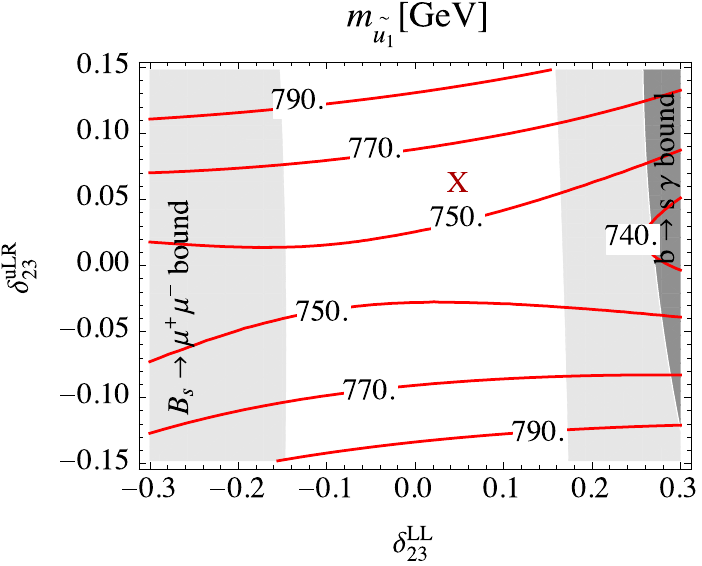}} \hspace*{-0.8cm}}}
   \label{fig2c}}
\caption{Dependence on the QFV parameters $\dll$ and $\dulr$ of the width (a) $\Gamma(h^0 \to c \bar{c})$ 
in MeV, (b) $\Gamma(h^0 \to c \bar{c})$/$\Gamma^{\rm SM}(h^0 \to c \bar{c})$ and (c) the mass of the lightest 
squark $\su_1$ in GeV. The light and dark gray regions are excluded by the constraints from the 
B($B_s \to \mu^+ \mu^-$) and B($b \to s \gamma$) data, respectively. 
}
\label{fig2}
\end{figure*}
%
%***********************************************************************
\begin{figure*}[h!]
\centering
\subfigure[]{
   { \mbox{\hspace*{-1cm} \resizebox{8.cm}{!}{\includegraphics{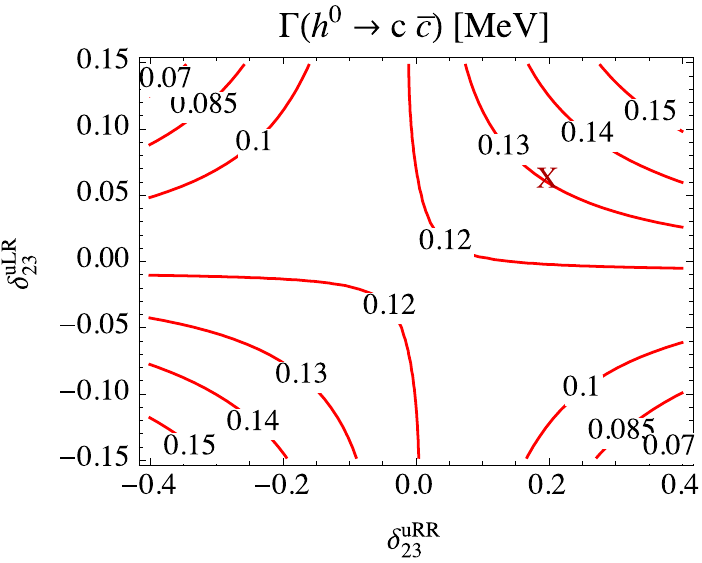}} \hspace*{-0.8cm}}}
   \label{fig3a}}
 \subfigure[]{
   { \mbox{\hspace*{+0.cm} \resizebox{8.cm}{!}{\includegraphics{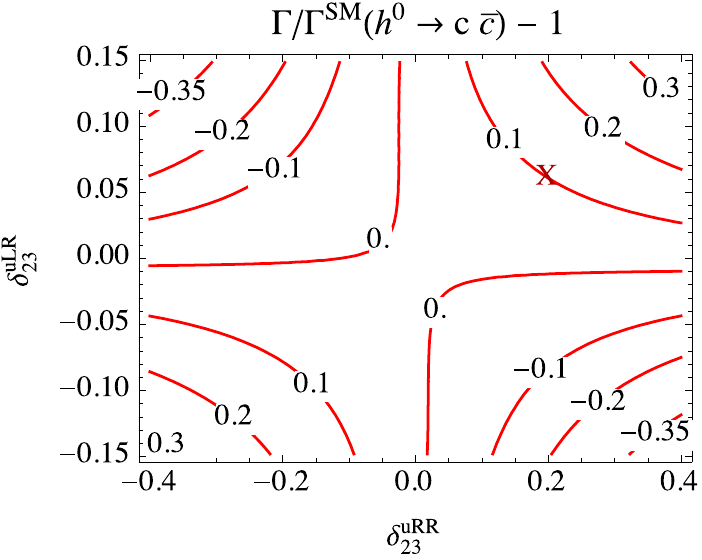}} \hspace*{-1cm}}}
  \label{fig3b}}\\
  \centering
  \subfigure[]{
   { \mbox{\hspace*{-1cm} \resizebox{8.cm}{!}{\includegraphics{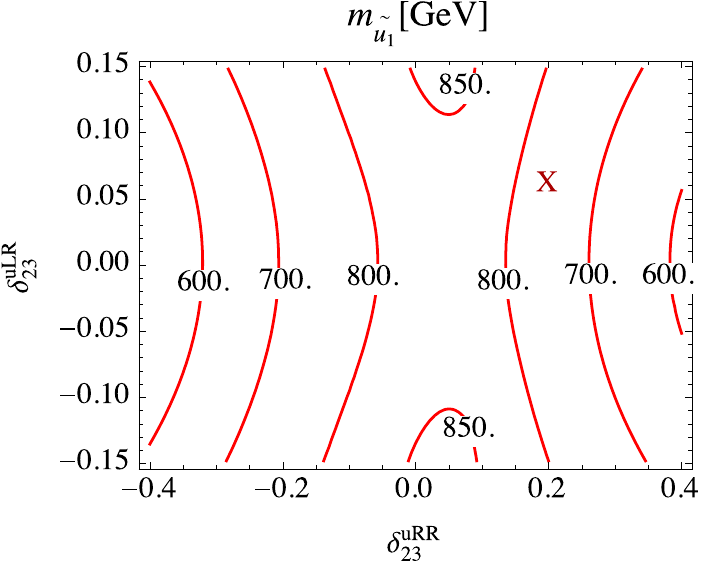}} \hspace*{-0.8cm}}}
   \label{fig3c}}
\caption{Dependence on the QFV parameters $\durr$ and $\dulr$ of the width (a) $\Gamma(h^0 \to c \bar{c})$ 
in MeV, (b) $\Gamma(h^0 \to c \bar{c})$/$\Gamma^{\rm SM}(h^0 \to c \bar{c})$ and (c) the mass of the lightest 
squark $\su_1$ in GeV.} 
\label{fig3}
\end{figure*}
%***********************************************************************
\begin{figure*}[h!]
\centering
\subfigure[]{
   { \mbox{\hspace*{-1cm} \resizebox{7.cm}{!}{\includegraphics{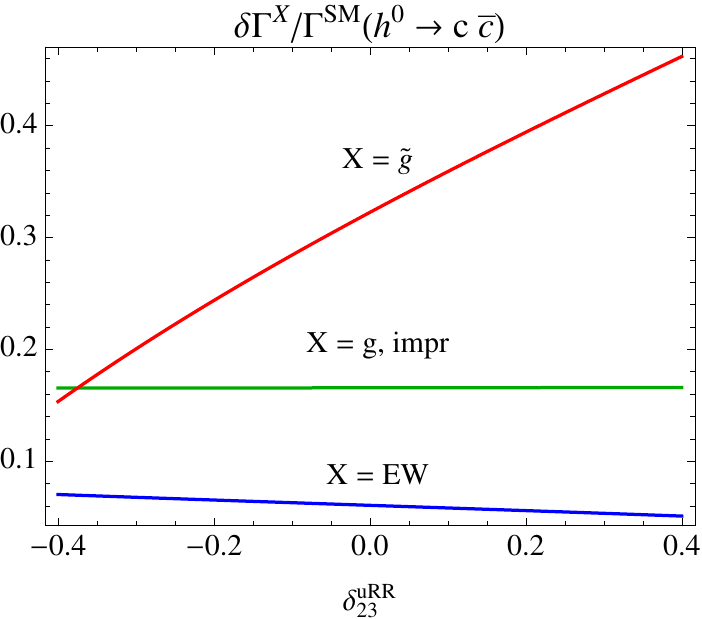}} \hspace*{-0.2cm}}}
   \label{contra}}
 \subfigure[]{
   { \mbox{\hspace*{+0.cm} \resizebox{7.cm}{!}{\includegraphics{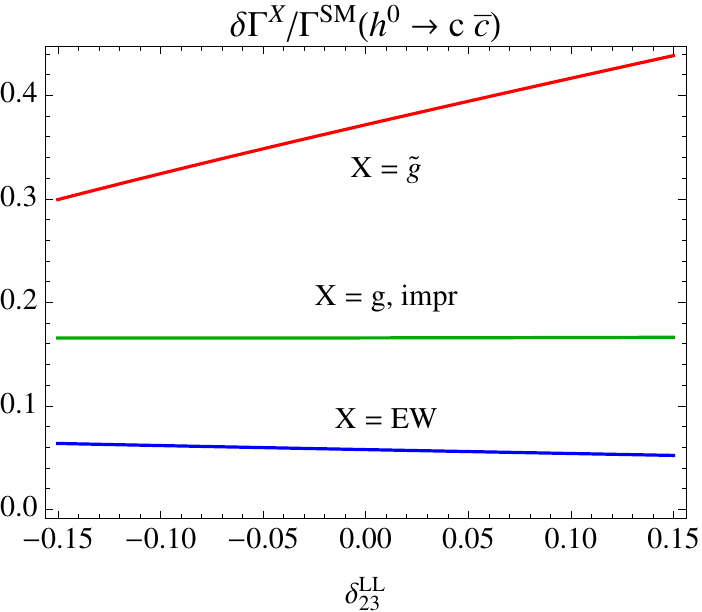}} \hspace*{-1cm}}}
  \label{contrb}}\\
  \centering
\subfigure[]{
   { \mbox{\hspace*{-1cm} \resizebox{7.cm}{!}{\includegraphics{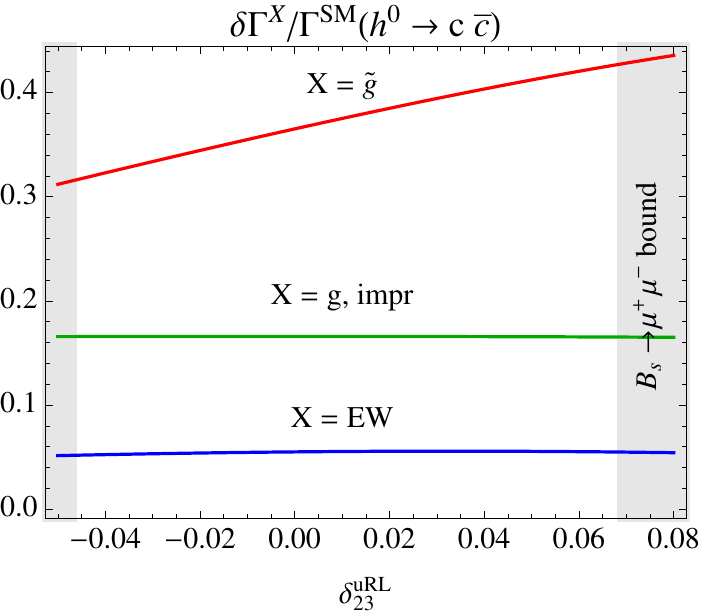}} \hspace*{-0.2cm}}}
   \label{contrc}}
 \subfigure[]{
   { \mbox{\hspace*{+0.cm} \resizebox{7.cm}{!}{\includegraphics{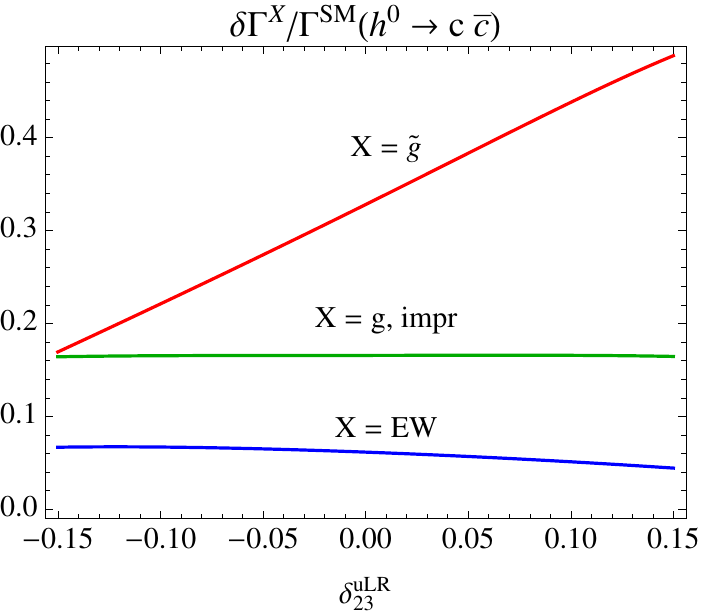}} \hspace*{-1cm}}}
  \label{contrd}}
\caption{Dependences on the QFV parameters of the one-loop $\sg$, EW and improved $g$ contributions to the width $\Gamma(h^0 \to c \bar{c})$. Note that for the $\sg$ and EW contributions only the one-loop scale independent part is shown as in the $\drbar$ scheme the scale dependent part cancels with the tree-level scale dependent part.} 
\label{contr}
\end{figure*}

The resulting physical masses of the particles are shown in Table~\ref{physmasses}. The flavor decomposition of the two lighter squarks $\su_1$ and $\su_2$ can be seen in Table~\ref{flavourdecomp}. This scenario satisfies all present experimental and theoretical constraints 
given in Appendix~\ref{sec:constr}. For calculating the masses and the mixing, as well as the low-energy observables, especially 
those in the B meson sector (see Table~\ref{TabConstraints}), we use the public code 
SPheno v3.3.3~\cite{SPheno1, SPheno2}. 
The width $\Gamma(h^0 \to c \bar{c})$ at full one-loop level in the MSSM with QFV is calculated on the basis of the formulas given above with the help of FeynArts~\cite{Hahn:2000kx} and FormCalc~\cite{Hahn:1998yk}. We also use the SSP package~\cite{SSP}.
In the following plots we show the QFV parameter dependences of the full 
one-loop level width $\Gamma(h^0 \to c \bar{c})$ of eq.~(\ref{fullres}) 
around the reference point of Table~\ref{basicparam}.

In Figs.~\ref{fig1a} and \ref{fig1b} we show the dependence of the width 
$\Gamma(h^0 \to c \bar{c})$ on the QFV parameters $\dll$ ($\ti{c}_L-\ti{t}_L$ mixing) 
and $\durr$ ($\ti{c}_R-\ti{t}_R$ mixing), with the other parameters fixed as in Table~\ref{basicparam}.
In Fig.~\ref{fig1a} we show the width in MeV as a function of $\dll$ and $\durr$.
The white area is the region allowed by all the constraints of
Appendix~\ref{sec:constr}, with the reference point of Table~\ref{basicparam} indicated by X. 
In the allowed region this width can vary from 0.1 MeV to 0.14 MeV. As can be seen, there is a rather strong dependence on  
$\durr$. 

In Fig.~\ref{fig1b}  we show the deviation of the $\Gamma(h^0 \to c \bar{c})$ 
from the SM width $\Gamma^{\rm SM}(h^0 \to c \bar{c})= 0.118$ MeV~\cite{pdg2014}. 
This deviation varies between -15\% and 20\%. It is interesting to mention that we
obtain $\Gamma^{\rm QFC}(h^0 \to c \bar{c})= 0.116$ MeV for the full one-loop width 
in the QFC MSSM case for our reference scenario corresponding to Table~\ref{basicparam}. This means that the QFC supersymmetric contributions 
change the width $\Gamma(h^0 \to c \bar{c})$ by only $\sim $ -1.5\% compared to 
the SM value.
Comparing our QFC one-loop result with FeynHiggs-2.10.2~\cite{Heinemeyer:1998yj} we have a difference less than 1\%. 
Note that the mass of the lightest squark $\su_1$ can vary in the allowed region between 650 GeV and 850 GeV, as seen in Fig.~\ref{fig1c}. 
Note also that in Figs.~\ref{fig1a} and \ref{fig1b} the QFV parameter $-0.3<\durr<0.3$ is not restricted by the constraints from the B sector, but from the mass of the lightest stop (corresponding to the lightest squark mass shown in Fig.~\ref{fig1c}) and the lightest neutralino (see Table~\ref{physmasses}) in the context of simplified MSSM with QFC 
~\cite{SUSY@ICHEP2014}. In principle, this experimental restriction on the lightest stop mass
does not hold for the case of QFV, and a wider range of $\durr$ is
allowed~\cite{Blanke}.

In Figs.~\ref{fig2a} and \ref{fig2b} we show the dependence of the width 
$\Gamma(h^0 \to c \bar{c})$ on the QFV parameters $\dll$ and $\dulr$ 
($\tilde{c}_L-\st_R$ mixing) with the other parameters fixed as in Table~\ref{basicparam}. In the allowed range the width can vary between 0.08 MeV and 0.15 MeV. The deviation of $\Gamma(h^0 \to c \bar{c})$ from the SM value $\Gamma^{\rm SM}(h^0 \to c \bar{c})$ lies between -30\% and 25\% (Fig.~\ref{fig2b}). Fig.~\ref{fig2c} shows the dependence of the mass $\msu{1}$.

In analogy we show in Fig.~\ref{fig3} the corresponding plots for the dependences on the QFV parameters $\durr$ and $\dulr$. As seen in Fig.~\ref{fig3a}, the width $\Gamma(h^0 \to c \bar{c})$ varies in the allowed region between 0.07 MeV and 0.15 MeV. The deviation from the SM value $\Gamma^{\rm SM}(h^0 \to c \bar{c})$ is between -35\% and 30\% (see Fig.~\ref{fig3b}). The mass of $\su_1$ varies between 600 GeV and 850 GeV, as seen in Fig.~\ref{fig3c}.

In Fig.~\ref{contr} we show the dependence of $\d \Gamma^X/\Gamma^{\rm SM}(h^0 \to c \bar{c})$ on the QFV 
parameters $\durr, \dll, \dulr$ and $\durl$ for the reference scenario of Table~\ref{basicparam}, where $\d \Gamma^X$ denotes the individual 
contribution of $X = (g, {\rm impr}), \sg, {\rm EW}$ (including the EW MSSM contributions) to the width 
$\Gamma(h^0 \to c \bar{c})$ (see eq.(\ref{fullres})). 
As can be seen, the gluino loop contribution $\d \Gamma^{\sg}$ depends significantly on $\durr$ 
and $\dulr$ with the dependences on $\dll$ and $\durl$ being somewhat weaker. 
The gluino loop contribution $\d \Gamma^{\sg}/\Gamma^{\rm SM}$ can go up to 45\% 
(see Figs.~\ref{contra} and \ref{contrd}). It can also be seen that the electroweak loop 
contributions $\d \Gamma^{\rm EW}$ cannot be neglected with $\d \Gamma^{\rm EW}/\Gamma^{\rm SM}$ 
being around 5\%. Clearly, its dependence on the QFV parameters is weak.

The strong dependences of the width $\Gamma(h^0 \to c \bar{c})$ on the QFV parameters 
shown in this section can be explained as follows.
First of all, the scenario chosen is characterized by large QFV parameters, which in our case are the large $\ti{c}_{L,R}-\ti{t}_{L,R}$ 
mixing parameters $\dll, \durr, \durl , \dulr$, 
and particularly large QFV trilinear couplings $T_{U23}, T_{U32}$ (Note that  $\durl \sim T_{U23}$ and $\dulr \sim T_{U32}$).    
In such a scenario, the lightest up-type squarks $\su_{1,2}$ are strong 
    admixtures of $\ti{c}_{L,R}-\ti{t}_{L,R}$, and, hence, the couplings 
    $\su_{1,2} \su_{1,2}^* h^0(\sim {\rm Re}(H_2^0))$ in Fig.~\ref{sgvertex} are strongly 
    enhanced, see eq.~(\ref{coupluiujh}).
In addition, large $\ti{t}_{L}-\ti{t}_{R}$ mixing due to the large QFC trilinear coupling $T_{U33}$ occurs. 
Moreover, the $\ti{t}_{L} \ti{t}_{L}^* h^0$ and  $\ti{t}_{R} \ti{t}_{R}^* h^0$ couplings are proportional 
to the top quark mass squared (see eq.~(\ref{coupluiujh})), which additionally  
enhances the $\su_{1,2} \su_{1,2}^* h^0$ couplings and thus also the vertex gluino contributions of Fig.~\ref{sgvertex} in case of QFV.

%\clearpage

%************************************************
\section{Observability of the deviation of  $\Gamma(h^0 \to c \bar{c})$ from its SM value at the ILC}
\label{sec:uncert}
%************************************************

Observation of any significant deviation of the width $\Gamma(h^0 \to c \bar{c})$ from its SM prediction 
signals new physics beyond the SM. It is important to estimate the 
uncertainties of the SM prediction reliably in order to confirm such a deviation. 
Once the deviation is discovered, one has to work out the new 
physics candidates suggesting it. 
 %***********************************************************************
\begin{figure*}[h!]
\centering
{ \mbox{\hspace*{-1cm} \resizebox{8.cm}{!}{\includegraphics{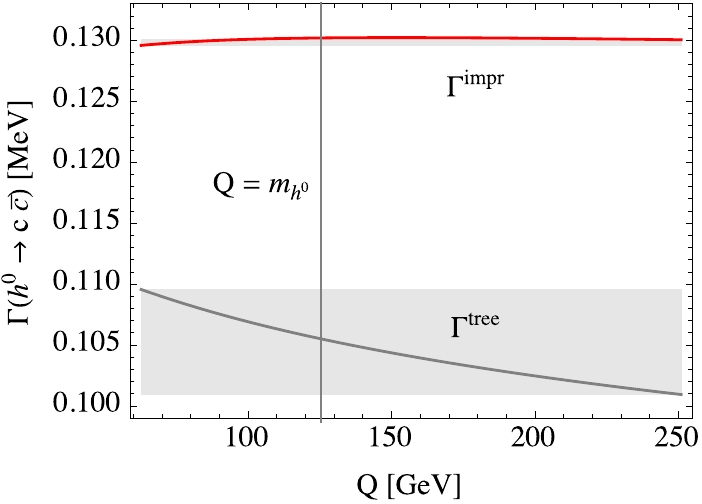}} \hspace*{-0.8cm}}}
\caption{Renormalization-scale dependence of $\Gamma(h^0 \to c \bar{c})$. $\Gamma^{\rm impr}(h^0 \to c \bar{c})$ is the improved one-loop corrected width of eq.~(\ref{fullres}). The vertical line shows ${\rm Q}=m_{h^0}$. }
\label{qdep}
\end{figure*}
 
The uncertainties of the SM prediction come from two sources~\cite{Dawson:2013bba, Heinemeyer:2013tqa, Almeida:2013jfa, Lepage:2014fla}. One is the
parametric uncertainty and the other is the theory uncertainty. The former is 
due to the errors of the SM input parameters such as $m_c(m_c)\vert_{\overline{\rm MS}}$ and 
$\alpha_s(m_Z)\vert_{\overline{\rm MS}}$, and the latter is due to unknown higher order corrections. 
The theory uncertainty is estimated mainly by renormalization-scale dependence 
uncertainties which are indicative of not knowing higher order terms in a 
perturbative expansion of the corresponding observable. These scale dependence uncertainties 
are estimated by varying the scale Q from ${\rm Q}/2$ to $2{\rm Q}$~\cite{Dawson:2013bba, Heinemeyer:2013tqa, Almeida:2013jfa}. (Note that in our case ${\rm Q}=m_{h^0}$.)

In order to estimate the uncertainty of the width $\Gamma(h^0 \to c \bar{c})$ in the MSSM with QFV at our reference point we proceed in an analogous way.
We calculate the parametric uncertainty in the width $\Gamma(h^0 \to c \bar{c})$ due to errors in the inputs $m_c (m_c)\vert_{\msb}$ and $\a_s (m_Z)\vert_{\msb}$ following~\cite{Yokoyatalk}
\be
\frac{\d \Gamma}{\Gamma}=\left|\frac{m_c}{\Gamma}\frac{\partial \Gamma}{\partial m_c}\right|\frac{\d m_c}{m_c} \oplus \left |\frac{\a_s}{\Gamma}\frac{\partial \Gamma}{\partial \a_s}\right|\frac{\d \a_s}{\a_s} \,,
\label{peskin}
\ee
where as input we take $m_c(m_c)\vert_{\overline{\rm MS}} = 1.275~\gev$ with $\d m_c/m_c=2\%$~\cite{PDG2013}, and $\a_s (m_Z)\vert_{\overline{\rm MS}} =0.1185$ with $\d \a_s/\a_s=0.5\%$~\cite{a_s@ICHEP2014}. $\d \rm X/X$ denotes the relative error of the quantity $\rm X$.
At our reference point of Table~\ref{basicparam} we get
\be
\frac{\d \Gamma}{\Gamma}=\vert2.6\vert \frac{\d m_c}{m_c} \oplus \vert-4.0\vert\frac{\d \a_s}{\a_s} =5.2\% \oplus 2\%\,.
\label{}
\ee
Note that the parametric uncertainties due to errors of the other SM input parameters, such as $m_b$, are negligible.

The theory uncertainty of the width for our reference point is shown on Fig.~\ref{qdep}. We have: $\d \Gamma/\Gamma(h^0 \to c \bar{c})=\scriptsize{\begin{array}{c} +0.11\%  \\ -0.46\% \end{array}}$, where $\Gamma(h^0 \to c \bar{c})$ is the improved one-loop corrected width of eq.~(\ref{fullres}). Thus, for this uncertainty we take $\sim 0.5\%$.

For the total error in the width at our reference point we get
\bea
\sqrt{5.2\%^2+2\%^2}+0.5\%&\approx& 6.1\%\,,
\label{ouruncert}
\eea
where the parametric uncertainties are added quadratically and the theory uncertainty is added to them linearly.
The obtained total uncertainty (\ref{ouruncert}) at our reference point is $\sim \pm $6.1\% (at 68\% CL), which is in good agreement with the estimated total uncertainty of $\Gamma^{\rm SM}(h^0 \to c \bar{c})$, see Table 13 of~\cite{Almeida:2013jfa}. Note that the uncertainty in the coupling is half of the uncertainty in the width.

As seen in Section~\ref{sec:num}, the deviation $\Gamma(h^0 \to c \bar{c})/\Gamma^{\rm SM}(h^0 \to c \bar{c})$ can be as large as $\sim \pm35$\%. 
Such a large deviation can be observed at ILC (500 GeV) with 1600 (500) ${\rm fb}^{-1}$, 
where the expected experimental error in the width is $\sim$3\% (5.6\%)~\cite{Asner:2013psa, Tian@EPS-HEP2013} .
A measurement of $\Gamma(h^0 \to c \bar{c})$ at LHC (even with the high luminosity upgrade) is demanding due to uncertainties in the charm-tagging.

%%%%%%%%%%%%%%%%
\section{Conclusions}
\label{sec:concl}
%%%%%%%%%%%%%%%%

We have calculated the width $\Gamma (h^0 \to c \bar{c})$  at full one-loop level within the MSSM with quark flavor violation. In particular, we have studied $\ti{c}_{R,L}-\ti{t}_{R,L}$ mixing, taking into account the experimental constraints from B-physics, $m_{h^0}$ and SUSY particle searches. The width $\Gamma (h \to c \bar{c})$ turns out to be very sensitive to $\ti{c}_{R,L}-\ti{t}_{R,L}$ mixing.

In our calculation we have used the $\drbar$ renormalization scheme. In particular, we have derived the explicit formula for the dominant gluino loop contribution. We also have performed a detailed numerical study of the QFV 
parameter dependence of the width. Whereas the width $\Gamma (h^0 \to c \bar{c})$ 
in the QFC MSSM case is only slightly different from its SM value, in the QFV 
case this width can deviate from the SM by up to $\sim \pm 35\%$.

We have estimated the theoretical uncertainties of 
$\Gamma (h^0 \to c \bar{c})$ and have shown that the SUSY QFV contribution 
to this width can be observed at the ILC.

%
%%%%%%%%%%%%%%%%%%%%%%%%%%%%%%%%%%%%%%%%%%%%%%%%%%%%%%%%%%%%%%%%%%%%%%%%%%%%
\section*{Acknowledgments}
%%%%%%%%%%%%%%%%%%%%%%%%%%%%%%%%%%%%%%%%%%%%%%%%%%%%%%%%%%%%%%%%%%%%%%%%%%%

We would like to thank W. Porod for 
helpful discussions, especially for the
permanent support concerning SPheno. K. H. thanks Prof. Koji Hashimoto for the warm hospitality at RIKEN Nishina Center 
for Accelerator-Based Science. This work is supported by the "Fonds zur F\"orderung der
wissenschaftlichen Forschung (FWF)" of Austria, project No. P26338-N27.

%Appendix
%%%%%%%%%%%%%%%%%%%%%%%%

\begin{appendix}

%%%%%%%%%%
\section{Interaction Lagrangian}
\label{sec:lag}
%%%%%%%%%

\begin{itemize}

\item The interaction of the lightest neutral Higgs boson, $h^0$, with two charm quarks is given by
\be
{\cal L}_{h^0 c \bar{c}}= s_1^c h^0 c \bar{c}\,,
\label{treelag}
\ee
where the tree-level coupling $s_1^c$ is given by eq.~(\ref{treecoup}).

\item  In the super-CKM basis, the interaction of the 
lightest neutral Higgs boson, $h^0$, with two up-type squarks is given by
\be
{\cal L}_{h^0\su_i \su_j}=G_{ij1}^{\su} h^0 \su_i^* \su_j, \, ~ i,j=1,...,6.
\ee
The coupling $G_{ij1}^{\su}$ reads
\begin{eqnarray}
&&  G_{ij1}^{\su}=-\frac{g} {2 m_W}\,
	\bigg[ -m_W^2 \sin(\alpha+\beta) \Big[
      (1-\tfrac13 \tan^2\thw) \nonumber \\
    &&  \times (U^{\su})_{jk} (U^{\su *})_{ik} + \tfrac43 \tan^2\thw (U^{\su})_{j\,(k+3)} 
    (U^{\su *})_{i\,(k+3)}
      \Big] \nonumber \\[0.2cm]
    & & +\ 2 \dfrac{\cos\alpha}{\sin\beta} \Big[
      (U^{\su})_{jk}\ m^2_{u,k} (U^{\su *})_{ik} 
     %&&
      + (U^{\su})_{j\,(k+3)} m^2_{u,k} (U^{\su *})_{i\,(k+3)}
      \Big] \nonumber \\[0.2cm]
    & & +\ \dfrac{\sin\alpha}{\sin\beta} \Big[
      \mu^* (U^{\su})_{j\,(k+3)} m_{u,k} (U^{\su *})_{ik} 
      %&&
      + \mu (U^{\su})_{jk} m_{u,k} (U^{\su *})_{i\,(k+3)}
      \Big] \nonumber \\[0.2cm]
    & & +\ \dfrac{\cos\alpha}{\sin\beta}\, \dfrac{v_2}{\sqrt2} \Big[
      (U^{\su})_{j(k+3)}\ (T_U)_{kl}\ (U^{\su *})_{il} 
    %&&
      + (U^{\su})_{jk}\ (T_U^*)_{lk}\ (U^{\su *})_{i\,(l+3)}
    \Big] \bigg] \, ,
    \label{coupluiujh}
\end{eqnarray}
where the sum over $k,l=1,2,3$ is understood.
Here $U^{\su}$ is the mixing matrix of the up-type squarks
\begin{eqnarray}
&&\su_{iL} = (U^{\su \dagger})_{ik} \su_k, \, \nonumber \\ 
&&\su_{iR} = (U^{\su \dagger})_{(i + 3)\,k} \su_k, \, ~ i=1,2,3,~ k=1,...,6.
\end{eqnarray}
Note that $(T_U)_{kl}$ in (\ref{coupluiujh}) are given in the SUSY Les Houche Accord notation\cite{Skands:2003cj}.

\item  The interaction of gluino, up-type squark and a charm quark is described by
\bea
{\cal L}_{\sg \su_i c}&=& -\sqrt{2} g_s T_{rs}^{\a}\bigg[\bar{\sg}^{\a}(U^{\su}_{i2}e^{-i\frac{\phi_3}{2}}P_L-U^{\su}_{i5}e^{i\frac{\phi_3}{2}}P_R) c^s \su_i^{*, r} \nn
&&+\bar{c}^r
(U^{\su *}_{i2} e^{i\frac{\phi_3}{2}}P_R-U^{\su *}_{i5}e^{-i\frac{\phi_3}{2}} P_L) \sg^{\a} \su_i^s\bigg]\,,
\eea
where $T^\a$ are the SU(3) colour group generators and summation over $r,s=1,2,3$ and over $\a=1,...,8$ is understood. In our case the parameter $M_3=\msg e^{i \phi_3}$ is taken as real, $\phi_3=0$.
\end{itemize}

%**********************************************************************
\section{Hard gluon/photon bremsstrahlung
\label{sec:brems}}
%**********************************************************************

The convergent one-loop gluon/photon corrected decay width in the limit of vanishing gluon/photon mass, $\lambda = 0$, is given by
\be
\Gamma^{g/\gamma}(h^0 \to c \bar{c})=\Gamma^{\rm tree} + \d \Gamma^{g/\gamma}+\Gamma^{\rm hard}(h^0 \to c \bar{c} g/\gamma)\,.
\ee
 The hard gluon radiation width reads
\be
\Gamma^{\rm hard}(h^0 \to c \bar{c} g)=\frac{2 \a_s |s_1^c|^2}{\pi^2 m_{h^0}}\left[ J_1-(m_{h^0}^2-4 m_c^2)(J_2-(m_{h^0}^2-2 m_c^2)J_3)\right]\,,
\label{hgluon}
\ee
with the integrals~\cite{Denner}
\be
J_1=\frac{1}{8 m_{h^0}^2} \left( (\kappa^2+6 m_c^4) \ln{\b_0}-\frac{3}{2}\kappa (m_{h^0}^2-2 m_c^2)\right)\,,
\ee
%**** 
\be
J_2=\frac{1}{4 m_{h^0}^2} \left( 2 \kappa \ln(\frac{\kappa^2}{\lambda m_{h^0} m_c^2}) -4 \kappa-m_{h^0}^2\ln{\b_0}\right)\,,
\ee
%**** 
\be
J_3=\frac{1}{2 m_{h^0}^2}\left(- \ln(\frac{\lambda m_{h^0} m_c^2}{\kappa^2})\ln{\b_0}+\ln^2{\b_0}-\ln^2{\b_1}+{\rm Li}_2(1-\b_0^2)-{\rm Li}_2(1-\b_1^2) \right)\,,
\ee
where ${\rm Li}_s (z)$ is the polylogarithm function, defined by the infinite sum
\be
{\rm Li}_s (z)=\sum_{k=1}^{\infty}\frac{z^k}{k^s}\,,
\label{li2}
\ee
 \be
 \b_0=\frac{m_{h^0}^2-2m_c^2+ \kappa}{2m_c^2}\,, \qquad \b_1=\frac{m_{h^0}^2- \kappa}{2m_{h^0} m_c}\,,
 \ee
 \be
\kappa\equiv \kappa (m_{h^0}^2, m_c^2,m_c^2)=m_{h^0}\sqrt{1-\frac{4 m_c^2}{m_{h^0}^2}}\,.
 \ee
 The expression for the hard photon radiation width $\Gamma^{\rm hard}(h^0 \to c \bar{c} \gamma)$ is obtained from (\ref{hgluon}) by making the replacements $C_F =4/3 \to e_c^2=4/9$ and $\a_s \to \a=e^2/(4 \pi)$.

%**********************************************************************
\section{Simplified formulas for the two- and three-point functions}
\label{app:PVsimpl}
%**********************************************************************

In our analytic calculations we neglect the squared masses of the charm quark and the lightest neutral Higgs boson, $m_c^2$ and $m_{h^0}^2$, in comparison to the squared masses of the scalar quarks and the gluino, $m_{\tilde{q}_i}^2$ and $\msg^2$. 
In the following we list the simplified expressions for the two- and three-point functions for this case.  
\be
 B_{0}(0, m_1^2, m_2^2)=\Delta +1+\frac{m_2^2 \ln \frac{m_2^2}{Q^2}-m_1^2 \ln \frac{m_1^2}{Q^2}}{m_1^2-m_2^2}
 \ee
\bea
 B_{1}(0, m_1^2, m_2^2)=-\frac{\Delta}{2}+\frac{1}{2}\left(\ln \frac{m_2^2}{Q^2}-\frac{m_1^4}{(m_1^2-m_2^2)^2}\ln \frac{m_2^2}{m_1^2}+\frac{m_2^2-3 m_1^2}{2(m_1^2-m_2^2)}\right)
\nonumber
\eea
\bea
=-\frac{\Delta}{2}+\frac{1}{4}\left(\ln \frac{m_1^2}{Q^2}+\ln \frac{m_2^2}{Q^2}-\frac{m_1^4+2m_1^2 m_2^2-m_2^4}{(m_1^2-m_2^2)^2} \ln \frac{m_2^2}{m_1^2} +\frac{m_2^2-3 m_1^2}{m_1^2-m_2^2}\right)
\eea
\be
 B_{0}(0, m^2,0)=\Delta +1- \ln \frac{m^2}{Q^2} 
 \label{B0simpl}
 \ee
 \be
 B_{1}(0, m^2,0)=\frac{-\Delta+\ln \frac{m^2}{Q^2}}{2}-\frac{3}{4} 
  \label{B1simpl}
 \ee
\be
 B_{0}(0, m^2,m^2)= B_{0}(0, m^2,0)-1=\Delta- \ln \frac{m^2}{Q^2} 
  \ee
 \be
 B_{1}(0, m^2,m^2)=-\frac{1}{2}B_{0}(0, m^2,m^2)
 \ee
  \be
 B_{0}(m^2,0, m^2)=\Delta +2+\ln \frac{Q^2}{m^2}
  \label{b0m0m}
 \ee
 \be
B_{1}(m^2,0, m^2)=-\frac{1}{2}\left(\Delta +1+\ln \frac{Q^2}{m^2} \right)
\label{b1m0m}
 \ee%
with $\Delta$ the UV divergence factor and $Q$ the renormalization scale.
%%%%%%%%%%%%%%%%%%%%%%%%%
\be
   \dot{B}_{0}(0, m_1^2, m_2^2)=\frac{m_1^4-m_2^4+2 m_2^2 m_1^2 \ln \frac{m_2^2}{m_1^2}}{2 \left(m_1^2-m_2^2\right){}^3}
\ee 
\be
   \dot{B}_{1}(0, m_1^2, m_2^2)=-\frac{2 m_1^6+3 m_2^2 m_1^4-6 m_2^4 m_1^2+m_2^6+6 m_2^2 m_1^4 \ln \frac{m_2^2}{m_1^2} }{6
   \left(m_1^2-m_2^2\right){}^4}
\ee   
\be
\dot{ B}_{0}(0, m^2,m^2)= \frac{1}{6 m^2} 
 \ee
\be
\dot{ B}_{0}(0, m^2,0)= \frac{1}{2 m^2} 
\label{dB0simpl}
 \ee
 \be
\dot{ B}_{0}( m^2,\lambda^2,m^2)= -\frac{1}{2m^2}\left(2-\ln \frac{m^2} {\lambda^2} \right)
 \ee
 \be
\dot{ B}_{1}( m^2,\lambda^2,m^2)= -\frac{1}{2m^2}
 \ee
 %
%
%***************************
\bea
{\rm Re}( C_{0}(m_1^2, m_2^2, , m_1^2,\lambda^2, m_1^2, m_1^2))&=& \frac{1}{m_{2}^2 \b}\bigg[ \ln \frac{1+\b}{1-\b}  \ln \frac{m_{2}^2 \b}{\lambda^2}-\frac{2 \pi^2}{3} \nonumber \\
&& \hspace*{-3cm } -2 {\rm Li}_2 \left(\frac{1-\b}{1+\b}\right)- \frac{1}{2}\ln^2 \left(\frac{1-\b}{1+\b}\right)
%2 \ln^2 \left(\frac{1+\b}{1-\b}\right)^{1/2}
+\ln \b \ln \frac{1+\b}{1-\b} \bigg]
\eea
\bea
{\rm Re}( C_{1}(m_1^2, m_2^2, , m_1^2,\lambda^2, m_1^2, m_1^2))&=&  -\frac{1}{m_{2}^2 \b} \ln \frac{1+\b}{1-\b} 
\eea
where $\b = (1-4m_1^2/m_2^2)^{1/2}$ and ${\rm Li}_s (z)$ is defined with (\ref{li2}).
\bea
C_{0}(0,0,0, m_1^2, m_2^2, m_3^2)&=&\frac{B_{0}(0, m_1^2, m_3^2)-B_{0}(0, m_2^2, m_3^2)}{m_1^2-m_2^2} \nonumber \\
&=&\frac{m_1^2 m_2^2 \ln \frac{m_1^2}{m_2^2}+m_2^2 m_3^2  \ln \frac{m_2^2}{m_3^2}+
  m_3^2 m_1^2 \ln \frac{m_3^2}{m_1^2} }{\left(m_1^2-m_2^2\right) \left(m_2^2-m_3^2\right)
   \left(m_1^2-m_3^2\right)}
\eea
\be
C_{0}(0,0,0, m_1^2, m_2^2, m_2^2)=\frac{m_1^2-m_2^2+m_1^2 \ln \frac{m_2^2}{m_1^2}}{\left(m_1^2-m_2^2\right){}^2}
\label{}
\ee
For $m_3=m_2 \ll m_1$ we get
\be
C_{0}(0,0,0, m_1^2, m_2^2, m_2^2)=\frac{1+ \ln \frac{m_2^2}{m_1^2}}{m_1^2}=\frac{1}{m_1^2}- \frac{\ln m_1^2}{m_1^2}+ \frac{\ln m_2^2}{m_1^2}
\label{C0simplxy}
\ee
\be
C_{0}(0,0,0, m^2, m^2, m^2)=-\frac{1}{2 m^2}
\ee
Note, that the expression (\ref{C0simplxy}) vanishes for fixed $m_2$ and $m_1 \to \infty$.

%%%%%%%%%%%%%%%%%%%%%%%%%%%%%%%%%%%%%%%%%%%%%%%%%%%%%%%%%%%%%%%%%%%%%%%
\section{Theoretical and experimental constraints}
\label{sec:constr}
%%%%%%%%%%%%%%%%%%%%%%%%%%%%%%%%%%%%%%%%%%%%%%%%%%%%%%%%%%%%%%%%%%%%%
% ============================================================================

Here we summarize the experimental and theoretical constraints taken into 
account in the present paper. The constraints on the MSSM parameters from the 
B-physics experiments and from the Higgs boson measurement at LHC are shown in 
Table~\ref{TabConstraints}. 

The BaBar and Belle collaborations have reported a slight 
excess of B$(B\to D\, \tau\,\nu)$ and B$(B\to D^*\,\tau\,\nu)$ \cite{Lees:2012xj, 
Lees:2013uzd, BELLE_ICHEP2014}.
However, it has been argued in \cite{Crivellin:2012ye} that within the
MSSM this cannot be explained without being at the same time in conflict
with B$(B_u\to \tau\,\nu)$. 
Using the program SUSY\_FLAVOR \cite{Crivellin:2012jv} we have checked that 
in our MSSM scenarios no significant enhancement occurs for B$(B\to D\, \tau\,\nu)$.
However, as pointed out in \cite{Nierste:2008qe}, the theoretical 
predictions (in the SM and MSSM) on B$(B \to D\, l\, \nu)$ and B$(B \to D^*\, l\, \nu)$~$
(l = \tau, \mu, e)$ have potentially large theoretical uncertainties due to the 
theoretical assumptions on the form factors at the $B\,D\,W^+$ and $B\,D^*\,W^+$ 
vertices (also at the $B\,D\,H^+$  and $B\,D^*\,H^+$ vertices in the MSSM). Hence 
the constraints from these decays are unclear. Therefore, we do not take these 
constraints into account in our paper.

In \cite{Dedes_t2ch} the QFV decays $t\to q h$ with $q=u,c$, have been studied in the 
general MSSM with QFV. It is found that these decays cannot be visible at the current 
LHC runs due to the very small decay branching ratios $B(t\to q h)$.

For the mass of the Higgs boson $h^0$, taking the naive combination of the ATLAS and 
CMS measurements \cite{Aad:2014aba,CMSHiggs} 
$m_{h^0} = 125.15 \pm 0.24~\gev$ \cite{Ellis_LHCP2014} and adding the 
theoretical uncertainty of $\sim \pm 2~\gev$ ~\cite{Heinemeyer_2012}
linearly to the experimental uncertainty at 2 $\sigma$, 
we take $m_{h^0} = 125.15 \pm 2.48 ~\gev$.

%
%*************************
\begin{table*}[h!]
\footnotesize{
\caption{
Constraints on the MSSM parameters from the B-physics experiments
relevant mainly for the mixing between the second and the third generations of 
squarks and from the data on the $h^0$ mass. The fourth column shows constraints 
at $95 \%$ CL obtained by combining the experimental error quadratically
with the theoretical uncertainty, except for $m_{h^0}$.
}
\begin{center}
\begin{tabular}{|c|c|c|c|}
    \hline
    Observable & Exp.\ data & Theor.\ uncertainty & \ Constr.\ (95$\%$CL) \\
    \hline\hline
    &&&\\
    $\Delta M_{B_s}$ [ps$^{-1}$] & $17.768 \pm 0.024$ (68$\%$ CL)~\cite{DeltaMBs_LHCb2013} 
    & $\pm 3.3$ (95$\%$ CL)~\cite{DeltaMBs_Carena2006, Ball_2006} &
    $17.77 \pm 3.30$\\
    $10^4\times$B($b \to s \gamma)$ & $3.40 \pm 0.21$ (68$\%$ CL)~\cite{PDG2013} 
    & $\pm 0.23$ (68$\%$ CL)~\cite{Misiak_2006} &  $3.40\pm 0.61$\\
    $10^6\times$B($b \to s~l^+ l^-$)& $1.60 ~ ^{+0.48}_{-0.45}$ (68$\%$ CL)~\cite{bsll_BABAR_2014}
    & $\pm 0.11$ (68$\%$ CL)~\cite{Huber_2008} & $1.60 ~ ^{+0.97}_{-0.91}$\\
    $(l=e~{\rm or}~\mu)$ &&&\\
$10^9\times$B($B_s\to \mu^+\mu^-$) & $2.9 \pm 0.7$ (68$\%$CL)~\cite{Bsmumu_LHCb, Bsmumu_CMS, Bsmumu_ICHEP2014}
& $\pm0.23$  (68$\%$ CL)~\cite{Bsmumu_SM_Bobeth_2014} & $2.90 \pm 1.44$ \\
$10^4\times$B($B^+ \to \tau^+ \nu $) & $1.15 \pm 0.23$  (68$\%$
CL)~\cite{Btotaunu_LP2013, Btotaunu_Babar_2013, Btotaunu_Belle_2013} 
&$\pm0.29$  (68$\%$ CL)~\cite{Btotaunu_LP2013} & $1.15 \pm 0.73$\\
$ m_{h^0}$ [GeV] & $125.03 \pm 0.30~(68\%~ \rm{CL}) (\rm{CMS})$ \cite{CMSHiggs}, 
&&\\
&$125.36 \pm 0.41~(68\%~ \rm{CL})(\rm{ATLAS})$ \cite{Aad:2014aba}
 & $\pm 2$~\cite{Heinemeyer_2012} & $125.15 \pm 2.48$ \\
&&&\\
    \hline
\end{tabular}
\end{center}
\label{TabConstraints}}
\end{table*}
%*************************
%

In addition to these constraints we also require our scenarios to be  
consistent with the following experimental constraints: 

(i) The LHC limits on the squark and gluino masses (at 95\% CL) 
    ~\cite{SUSY@ICHEP2014, SUSY@ICHEP2012, 
    SUSY2@ICHEP2012, SUSY@LHC, Aad:2012ms, AtlasConf2012, Aad:2012hm, 
    SUSY1@LHC, SUSY2@LHC, Chatrchyan:2012rg, Chatrchyan:2012ira, Chatrchyan:2012qka, 
    stop_sbot@LHC, Aad:2012yr, Aad:2012tx, Aad:2012xqa, Aad:2012ywa, Aad:2012pq, 
    ATLAS:2012ah, ATLAS:2012ai, Aad:2011cw, CMS, Chatrchyan:2013lya}: \\
In the context of simplified models, gluino masses $\msg \lesssim 1~{\rm TeV}$ are 
excluded at 95\% CL. The mass limit varies in the range 1000-1400~GeV depending 
on assumptions. 
First and second generation squark masses are excluded below 900~GeV. 
Bottom squarks are excluded below 600~GeV.
A typical top-squark mass limit is $\sim$ 700~GeV. 
In~\cite{CMS,Chatrchyan:2013lya} a limit for the mass of the top-squark $m_{\tilde{t}} 
\gsim 500~\gev$ for $m_{\tilde{t}} - m_{\rm LSP}=200~\gev$ is quoted. 
Including mixing of $\sca_R$ and $\sto_R$ would even lower this limit~\cite{Blanke}. 

(ii) The LHC limits on $m_{\ch_1}$ and $m_{\nt_1}$ from negative
searches for charginos and neutralinos mainly in leptonic final states
~\cite{SUSY@ICHEP2014, chargino_neutralino@LHC, Chatrchyan:2012pka}.

(iii) The constraint on ($ m_{A^0, H^+} , \tan \b $) from the MSSM Higgs boson
searches at LHC 
      ~\cite{Aad:2014aba, Kado, CMSHiggs, David, MSSM_Higgs@LHC}.

(iv) The experimental limit on SUSY contributions on the electroweak
$\rho$ parameter
     ~\cite{Altarelli:1997et}: $\Delta \rho~ (\rm SUSY) < 0.0012.$

Furthermore, we impose the following theoretical constraints from the vacuum 
stability conditions for the trilinear coupling matrices~\cite{Casas}: 
\begin{eqnarray}
|T_{U\alpha\alpha}|^2 &<&
3~Y^2_{U\alpha}~(M^2_{Q \alpha\alpha}+M^2_{U\alpha\alpha}+m^2_2)~,
\label{eq:CCBfcU}\\[2mm]
|T_{D\alpha\alpha}|^2 &<&
3~Y^2_{D\alpha}~(M^2_{Q\alpha\alpha}+M^2_{D\alpha\alpha}+m^2_1)~,
\label{eq:CCBfcD}\\[2mm]
|T_{U\alpha\beta}|^2 &<&
Y^2_{U\gamma}~(M^2_{Q \beta\beta}+M^2_{U\alpha\alpha}+m^2_2)~,
\label{eq:CCBfvU}\\[2mm]
|T_{D\alpha\beta}|^2 &<&
Y^2_{D\gamma}~(M^2_{Q \beta\beta}+M^2_{D\alpha\alpha}+m^2_1)~,
\label{eq:CCBfvD}
\end{eqnarray}
where
$\a,\b=1,2,3,~\a\neq\b;~\gamma={\rm Max}(\a,\b)$ and
$m^2_1=(m^2_{H^+}+m^2_Z\sin^2\theta_W)\sin^2\b-\frac{1}{2}m_Z^2$,
$m^2_2=(m^2_{H^+}+$  
$m^2_Z\sin^2\theta_W)$ $\cos^2\beta-\frac{1}{2}m_Z^2$.
The Yukawa couplings of the up-type and down-type quarks are
$Y_{U\alpha}=\sqrt{2}m_{u_\alpha}/v_2=\frac{g}{\sqrt{2}}\frac{m_{u_\alpha}}{m_W
\sin\beta}$
$(u_\a=u,c,t)$ and
$Y_{D\alpha}=\sqrt{2}m_{d_\alpha}/v_1=\frac{g}{\sqrt{2}}\frac{m_{d_\alpha}}{m_W
\cos\beta}$
$(d_\a=d,s,b)$,
with $m_{u_\a}$ and $m_{d_\a}$ being the running quark masses at the weak
scale and $g$ being the SU(2) gauge coupling. All soft SUSY-breaking parameters 
are given at $\rm Q=125.5~\gev$. As SM parameters we take $m_Z=91.2~\gev$ and
the on-shell top-quark mass $m_t=173.3~\gev$ \cite{top_mass@ICHEP2014}.
We have found that our results shown are fairly
insensitive to the precise value of $m_t$.

\end{appendix}
%%%%%%%%%%%%%%%%%%%%%%%%%

%
%
%Bibliography %=============================
%\section*{References}

\end{document}